\documentclass[pre,onecolumn,showpacs,preprintnumbers,amsmath,%
amssymb,superscriptaddress,eqsecnum,showpacs,
   preprintnumbers,floatfix]{revtex4-1}[12pt]
\usepackage{color}
\usepackage{morefloats}
\usepackage{graphicx}
\usepackage{subfigure}

\begin{document}

\newcommand{\vphi}{\varphi}
\newcommand{\bq}{\begin{equation}}
\newcommand{\be}{\begin{equation}}
\newcommand{\ba}{\begin{eqnarray}}
\newcommand{\eq}{\end{equation}}
\newcommand{\ee}{\end{equation}}
\newcommand{\ea}{\end{eqnarray}}
\newcommand{\tchi} {{\tilde \chi}}
\newcommand{\tA} {{\tilde A}}
\newcommand{\sech} { {\rm sech}}
\newcommand{\pstar}{\mbox{$\psi^{\ast}$}}
\newcommand {\bPsi}{{\bar \Psi}}
\newcommand {\bpsi}{{\bar \psi}}
\newcommand {\barf}{{\bar f}}
\newcommand {\barv}{{\bar v}}
\newcommand {\tu} {{\tilde u}}
\newcommand {\tv} {{\tilde v}}
\newcommand{\dq}{{\dot q}}
\newcommand {\tdelta} {{\tilde \delta}}
 \newcommand{\shao}[1]{\textcolor{red}{#1}}
 \newcommand{\etc}{\textit{etc}{}}
 \newcommand{\ie}{\textit{i.e.}{~}}
 \newcommand{\videpost}{\textit{vide post}{}}
 \newcommand{\eg}{\textit{e.g.}{~}}
 \newcommand{\ansatz}{\textit{ansatz}{ }}
 \
\maxdeadcycles=1000

\preprint{LA-UR 16- (Draft fnlde2.tex) : \today}

\title{Nonlinear Dirac equation solitary waves under a spinor force with different components}
\author{Franz G.  Mertens}
\email{franzgmertens@gmail.com}
\affiliation{Physikalisches Institut, Universit\"at Bayreuth, D-95440 Bayreuth, Germany} 
\author{Fred Cooper}
\email{cooper@santafe.edu}
\affiliation{Santa Fe Institute, Santa Fe, NM 87501, USA}
\affiliation{Theoretical Division and Center for Nonlinear Studies, 
Los Alamos National Laboratory, Los Alamos, New Mexico 87545, USA}
\author{Sihong Shao}\email{sihong@math.pku.edu.cn}
\affiliation{LMAM and School of Mathematical Sciences, Peking University, Beijing 100871, China}
\author{Niurka R. Quintero} 
\email{niurka@us.es} 
\affiliation{IMUS and Departamento de Fisica Aplicada I, E.P.S. Universidad de Sevilla, E-41011 Sevilla, Spain}

\author{Avadh Saxena}
\email{avadh@lanl.gov}
\affiliation{Theoretical Division and Center for Nonlinear Studies, 
Los Alamos National Laboratory, Los Alamos, New Mexico 87545, USA}
\author{A.R.  Bishop}
\email{arb@lanl.gov}
\affiliation{Los Alamos National Laboratory, Los Alamos, New Mexico 87545, USA}

\begin{abstract}
We consider the nonlinear Dirac (NLD) equation in 1+1 dimension with 
scalar-scalar  self-interaction 
 in the presence of  external forces as well as damping of the form  $\gamma^0 f(x,t) - i \mu  \gamma^0 \Psi$, where both $f, \{f_j = r_i e^{i K_j x}  \}$ and $\Psi$ are two-component spinors. We develop an approximate variational approach using collective coordinates (CC) for studying the time dependent response of the solitary waves to these external forces.  In our previous paper we assumed $K_j=K, ~ j=1,2$ which allowed a transformation to a simplifying coordinate system, and we also assumed  the ``small'' component of the external force was zero.    Here we include the effects of the small component and also the case $K_1 \neq K_2$ which dramatically modifies the behavior of the solitary wave in the presence of these external forces.	
\end{abstract}

\maketitle

\section{Introduction}

Ever since the nonlinear generalization of the Dirac equation \cite{Ivanenko1938}, the nonlinear Dirac (NLD) equation  
has found many applications as a practical model in numerous physical systems, e.g. extended particles 
\cite{FinkelsteinLelevierRuderman1951, FinkelsteinFronsdalKaus1956,Heisenberg1957}, nonlinear optics \cite{Barashenkov1998}, 
waveguide arrays as well as experimental optical realization of relativistic quantum mechanics 
\cite{Longhi2010,DreisowHeinrichKeil2010,TranLonghiBiancalana2014},
and  honeycomb optical lattices harboring Bose-Einstein condensates \cite{Haddad2009}. 
The NLD equation also arises in the context of phenomenological models of quantum chromodynamics \cite{Fillion-Gourdeau2013} 
and the influence of matter on the evolution of the Universe in cosmology \cite{Saha2012}.
In order to keep  the Lorentz invariance of the NLD equation,
the self-interaction Lagrangian can be obtained from the bilinear covariants. Different  NLD equations results from different self-interactions.  
A variety of models have been proposed and explored using the 
scalar bilinear covariant \cite{Gursey1956,Soler1970,GrossNeveu1974,Mathieu1984},
the vector bilinear covariant \cite{Thirring1958} 
and the axial vector bilinear covariant \cite{Mathieu1983}.  Moreover, models involving 
both scalar and pseudoscalar bilinear covariants \cite{RanadaRanada1984} 
as well as both scalar and vector bilinear covariants \cite{Stubbe1986-jmp,NogamiToyama1992} were studied.


An important aspect of these NLD equations is that
they allow solitary wave solutions or particle-like solutions: localized solutions with finite energy and charge \cite{Ranada1983}.
In other words, the particles appear as intense localized regions of field which can be identifieed as the basic ingredient
in the description of extended objects in quantum field theory \cite{Weyl1950}. For the  (1+1) dimensional NLD equation (\ie one 
time dimension plus one space dimension), several analytical solitary wave solutions were derived 
for the quadratic nonlinearity \cite{LeeKuoGavrielides1975,ChangEllisLee1975}, 
for fractional nonlinearity \cite{Mathieu1985-prd} as well as 
for general nonlinearity \cite{Stubbe1986-jmp,CooperKhareMihailaSaxena2010,XuShaoTangWei2013}
by invoking explicitly the constraints arising from energy-momentum conservation; 
which is well summarized by Mathieu \cite{Mathieu1985-jpa-mg}.
Using the analytical expressions of the NLD solitary wave solutions, the interaction dynamics among them 
 has been investigated and rich nonlinear phenomena have been brought out in a series of works 
\cite{AlvarezCarreras1981,ShaoTang2005,ShaoTang2006,ShaoTang2008,XuShaoTang2013,TranNguyenDuong2014}.


The stability of the NLD equation solitary waves is an important topic, which has been studied for several decades.
There are serious difficulties with the analytical studies of the NLD solitary wave stability \cite{StraussVazquez1986,AlvarezSoler1986,BlanchardStubbeVazquez1987}. 
On the other hand,  simulations results seem to lead to contradictory results \cite{Bogolubsky1979-pla,AlvarezSoler1983,Mathieu1983,Alvarez1985}.
From the numerical results it follows that both the multi-hump  profile and high-order nonlinearity could
affect the stability during the scattering of the NLD solitary
waves \cite{ShaoTang2005,XuShaoTang2013}.  In the case of NLD equation with scalar-scalar interactions (the Soler model)  the solitary wave solutions can have either one hump or two humps.

Recently, for the Soler model,  we found that all stable NLD solitary waves have a one-hump profile, but not all one-hump waves are stable, while all waves with two humps are unstable \cite{ShaoQuinteroMertens2014}. This result is consistent with the rigorous analysis in the nonrelativistic limit \cite{ComechGuanGustafson2014}.  The spectral analysis of the NLD equation was 
also recently studied \cite{NabileComech2016}. 
For a better understanding of  the behavior and the stability of NLD solitary waves, the NLD equation in the presence of external potentials was also investigated \cite{NogamiToyamaZhao1995,Toyama1998,ToyamaNogami1998,MertensQuinteroCooper2012}
and a sufficient dynamical condition for instability was postulated through a collective coordinates (CC) theory \cite{MertensQuinteroCooper2012}. 
In this work, we will continue to study the NLD solitary waves under external forces.


For the forced nonlinear Schr{\"o}dinger (NLS) equation  when subject to 
an external force of the form $f(x)= r \exp(-i K x)$, the authors found \cite{MertensQuinteroBarashenkovBishop2011,QuinteroMertensBishop2015,MertensQuinteroBishop2010} 
that intrinsic soliton oscillations are excited, i.e., the soliton amplitude, width, phase, momentum, and velocity all oscillate with the same frequency. This behavior was predicted by a collective coordinates 
theory and was confirmed by numerical simulations. Moreover, one specific plane wave phonon (short for a linear excitation) with wavenumber $k=-K$ is also excited such that the total momentum in a transformed NLS equation is conserved. This phonon mode was not included in the CC theory and had 
to be calculated separately \cite{MertensQuinteroBishop2013}.


In the present paper we consider the relativistic generalization of our previous work on the forced NLS equation, namely the behavior of solitary wave solutions to the NLD equation when subjected to an external force which is now a two-component spinor. In Sec. \ref{sec2} 
we review exact analytical solutions for the unperturbed NLD equation. In Sec. \ref{sec3} 
we present the NLD equation with external force $f_{j}(x,t)=r_{j} \exp[i (\nu_j t-K_j x)]$, $j=1,2$, and the corresponding Lagrangian density. Using the energy-momentum tensor we show that the total energy is conserved if the force is time independent ($\nu_j=0$).

For the case $K_1=K_2=K$, $\nu_j=0$ and zero dissipation it was possible to perform a transformation such that the transformed NLD equation is invariant under space translations and thus the momentum was also conserved.  In that case, when we set $r_2=0$ also we showed in (I) \cite{us1}  that the collective coordinates approach for studying the behavior of the solitary waves under the influence of these external forces agreed well with numerical solutions of the exact equations.  Here we loosen the restriction on the $K_j$ and also allow $r_2 \neq 0$ with the caveat $|r_2 |<  |r_1| $. 
In Sec. \ref{sec5} we make a variational ansatz with three collective coordinates. All integrals that appear in the Lagrangian can be performed exactly and we finally have a set of two first-order ODEs and one second-order ODE as  our CC equations. This is to be contrasted with the special case $K_1=K_2$  considered  in $(I)$ \cite{us1} where 
the CC equations consisted of two first order ODEs and one constraint equation.

\section{Review of exact solutions to the  NLD equation} \label{sec2}
We first review the known exact solitary wave solutions to the NLD equation,
\bq
(i \gamma^{\mu} \partial_{\mu} - m) \Psi +g^2 (\bPsi  \Psi)^{\kappa} \Psi 
= 0 \>, \label{nlde1}
\eq
where we use the representation for the 1+1 dimensional Dirac Gamma matrices:
$\gamma^0 = \sigma_3$;  ~~~$\gamma^1= i \sigma_2,$ which we also used previously  \cite{MertensQuinteroCooper2012}. 
In the rest frame the solitary wave solution is represented by 
\bq \label{eq2.3}
\Psi(x,t) = e^{-i\omega t} \psi(x)= e^{-i\omega t} \left(\begin{array} {cc}
      A(x) \\
      i ~B(x) \\ 
   \end{array}\right), 
\eq
where $A$ and $B$ satisfy
\ba
&& \frac{dA}{dx} + (m+\omega ) B - g^2(A^2-B^2)^{ \kappa} B=0\,, \nonumber \\
&&\frac{dB}{dx} + (m-\omega ) A - g^2(A^2-B^2)^{ \kappa} A=0\,. \nonumber \\
\ea
The solutions of these equations that vanish at infinity are given by 
\ba \label{eq2.33}
A & = & \sqrt{ \frac{(m+\omega)  \cosh ^2(\kappa \beta x)}{m+\omega \cosh(2\kappa \beta x)}} 
\bigg [\frac{(\kappa+1) \beta ^2}{g^2 (m+\omega  \cosh (2\kappa \beta x))} 
\bigg ]^{\frac{1}{2\kappa}}, \nonumber   \\ 
B & =  & \sqrt{ \frac{(m-\omega) \sinh^2(\kappa \beta x)}{m+\omega \cosh(2\kappa \beta x)}} 
\bigg [\frac{(\kappa +1) \beta ^2}{g^2 (m+\omega  \cosh (2\kappa \beta x))} 
\bigg ]^{\frac{1}{2\kappa}}, \nonumber   \\ 
\ea
where $\beta = \sqrt{m^2-\omega^2}$.
We search for bound state solutions which have positive frequency $\omega > 0$ and energies in the rest frame that are smaller than the mass parameter $m$, i.e. $\omega < m$. 

  Invoking Lorentz invariance we can obtain the solution in a frame moving with velocity $v$ with respect to the rest frame.
  The Lorentz boost  is given in terms of  the rapidity variable $ \eta$ as follows  (setting $c=1$): 
  \bq
  v = \tanh \eta;~~   \gamma = \frac{1}{\sqrt{1-v^2}} = \cosh \eta; ~~ \sinh \eta =  \frac{v}{\sqrt{1-v^2}} . 
  \eq
  
  In the moving frame, the transformation law for spinors implies that:
  \bq
  \Psi(x,t) =    \left(\begin{array}{cc}
        \cosh(\eta/2) & \sinh(\eta/2) \\
        \sinh(\eta/2) &  \cosh(\eta/2\\
     \end{array} \right)  \left(  \begin{array} {cc}
      \Psi_1^0[\gamma(x-vt), \gamma(t-vx)] \\
  \Psi_2^0[\gamma(x-vt), \gamma(t-vx)]\\ 
   \end{array} \right) , 
\eq
since
\bq
\cosh (\eta/2) = \sqrt{(1+\gamma)/2};~~  \sinh (\eta/2) = \sqrt{(\gamma-1)/2} . 
 \eq
 In component form this reads:
 \ba \label{eq2.37}
 &&\Psi_1(x,t) = \left( \cosh(\eta/2) A(x') + i \sinh(\eta/2) B(x') \right) e^{-i\omega t'}  , \nonumber \\
&&\Psi_2 (x,t) = \left( \sinh(\eta/2) A(x') + i \cosh(\eta/2) B(x') \right) e^{-i\omega t'}  ,
\ea
where
\bq
x' = \gamma(x-vt); ~~ t' = \gamma(t-vx) . 
\eq
Note that  $\cosh^2(\eta/2)+\sinh^2(\eta/2) = \cosh \eta = \gamma$.

\section{Externally Driven NLD equation} \label{sec3}

In  previous papers \cite{MertensQuinteroBishop2010,MertensQuinteroBarashenkovBishop2011} we 
investigated the externally driven NLS equation 

\bq
 i \frac{\partial}{\partial t} \psi + \frac{\partial^2}{\partial x^2} \psi + {g ^2}(\psi^\star \psi)^{ \kappa} \psi+\delta \psi  = r e^{-i K x}-i\mu \psi  \label{psieq} , 
\eq
where $\mu$ is the dissipation coefficient, and $r$ and $K$ are constants.  
This equation can be derived by means of a generalization of the Euler-Lagrange 
equation  
\bq \label{df2}
\frac{d}{dt} \frac{\partial {\cal L}}{\partial \psi_{t}^{*}} + 
\frac{d}{dx} \frac{\partial {\cal L}}{\partial \psi_{x}^{*}}-
\frac{\partial {\cal L}}{\partial \psi^{*}}= 
\frac{\partial {\cal F} }{\partial \psi^{*}_{t}},
\eq
where the Lagrangian density reads  
\bq \label{df3}
{\cal L} = \frac{i}{2} (\psi_{t} \psi^{*}-\psi_{t}^{*} \psi)-|\psi_{x}|^{2}+ 
\frac{g^2} {\kappa+1} (\psi^\star \psi)^{\kappa+1}  +\delta |\psi|^{2}-r e^{-i K x} \psi^{*}- r e^{i K x}\psi,
\eq
and the dissipation function density  is given by 
\bq \label{df4}
{\cal F} = -i \mu (\psi_{t} \psi^{*}-\psi_{t}^{*} \psi).
\eq

For the NLD case we instead consider a two-component spinor forcing term
\bq \label{force0}
f = \left({\begin{array}{c} 
  f_1(x,t) \\ 
  f_2(x,t) \\ 
\end{array}} \right)
\eq
with the NLD equation
\bq
(i \gamma^{\mu} \partial_{\mu} - m) \Psi +g^2 (\bPsi  \Psi)^{\kappa} \Psi 
=   \gamma^0 f(x,t) - i \mu \gamma^0 \Psi  \>. \label{nlde1a}
\eq
In what follows we will generalize our choice for the NLS equation by choosing 
\bq \label{force}
f_j(x,t) =  r_j e^{i( \nu_j t - K_j x)}, \quad j=1,2, 
\eq
with real parameters $r_j$, $\nu_j$ and $K_j$. 
Note that the phase of $f$ is invariant under Lorentz transformations. 
As the second component of the spinor $\Psi$ is the so-called ``small 
component", which is smaller than the first component by the factor 
$\alpha=\sqrt{(m-\omega)/(m+\omega)}$, 
we will only consider cases with $r_2=\epsilon r_1$, where $\epsilon=O(\alpha)$ 
or smaller. 

Equation \ (\ref{nlde1a}) can be derived in a standard fashion from the Lagrangian density
\bq \label{eq3.8}
\mathcal{L} =  \left(\frac{i}{2}\right) [\bPsi \gamma^{\mu} \partial_{\mu} \Psi 
-\partial_{\mu} \bPsi \gamma^{\mu} \Psi] - m \bPsi \Psi 
+ \frac{g^2}{\kappa+1} (\bPsi \Psi)^{\kappa+1} - \bPsi  f - \bar{f}  \Psi + \mathcal{L}_0(b),
\eq
where $\mathcal{L}_0(b)$ is determined later on and $b=\lim_{x \to \pm \infty} \Psi(x,t)$. 
The term in the Lagrangian density which pertains to  forcing  can be written as 
\bq
{\cal L} _3 = - 2 Re (\barf \Psi),
\eq
and the full interaction part of the Lagrangian density is now
\bq
{\cal L}_I = \frac{g^2}{\kappa+1} (\bPsi \Psi)^{\kappa+1} - \bPsi  f - \bar{f}  \Psi  . \ 
\eq
The generalized Euler-Lagrange equation can be written as
\bq \label{E-L} 
\partial_\mu \frac{\partial \mathcal{L}}{ \partial( \partial_\mu \bPsi)} -  \frac{\partial \mathcal{L}}{\partial \bPsi} = \frac{\partial \mathcal{F}}{\partial(\partial_t \bPsi)} ,
\eq
where the dissipation function density is now
\bq \label{df5}
{\cal F} =-i \mu ( \bPsi \gamma^0 \partial_t \Psi - \partial_t \bPsi \gamma^0 \Psi ) .
\eq

The adjoint equation comes from the Euler-Lagrange equation:
\bq 
\partial_\mu \frac{\partial \mathcal{L}}{ \partial( \partial_\mu \Psi)} -  \frac{\partial \mathcal{L}}{\partial \Psi} = \frac{\partial \mathcal{F}}{\partial(\partial_t \Psi)},
\eq
from this we get the adjoint driven NLD equation
\bq
-i \partial_{\mu}  \bPsi \gamma^\mu - m \bPsi  +g^2 (\bPsi  \Psi)^{\kappa}  \bPsi 
=  \barf  \gamma^0 + i \mu \bPsi \gamma^0  \>.  \label{nldradj}
\eq
To generalize our discussion of external forces from the NLS equation to the NLD equation we have included a dissipation term in our general formulation. However, in most sections that follow we will concentrate on the case where the dissipation coefficient $\mu=0$, so that the energy is conserved. 
In (I) \cite{us1} we restricted our discussion to $K_1=K_2=K$ which led to conservation of momentum in a particular frame.  In this paper we lift this restriction.

\subsection{Energy flow equations and the conservation of energy}

From the NLD equation with external sources and the definition of the energy-momentum tensor:
\bq  \label{emc1}
 T^{\mu \nu} = \frac{i}{2} \left[ \bPsi \gamma^\mu \partial^ \nu \Psi  -  \partial^\nu \bPsi  \gamma^\mu  \Psi \right]  - g^{\mu \nu} \cal{L}, 
 \eq
we have that 
\bq
\partial_\mu T^{\mu \nu} = F^\nu, 
\eq
where
\bq \label{fdens}
F^\nu = \bPsi (\partial^\nu  f)+ (\partial^\nu  \barf )\Psi . 
\eq
The energy density is given by 
\bq 
T^{00} = -\frac{i}{2} \left[  \bPsi  \gamma^1 \partial_x \Psi-\partial_x \bPsi \gamma^1 \Psi \right]+ m \bPsi \Psi - \mathcal{L}_I - 
\mathcal{L}_0,    \label{eq:hdensity2}
\eq
where  now 
\bq
 \mathcal{L}_I  =   \frac{g^2}{ \kappa+1} (\bPsi \Psi)^{ \kappa+1} -\barf \Psi -\bPsi f, \ 
\eq
and $\mathcal{L}_0$ is chosen so that $T^{00}$ vanishes at $x= \pm L$, when $L \to \infty$. Therefore, 
from Eqs. (\ref{eq3.8}) and (\ref{eq:hdensity2}) we obtain  
\bq
 \mathcal{L}_0  = m \bar b  b - \frac{g^2}{ \kappa+1} (\bar b  b)^{ \kappa+1} + \bar b r  + \bar{r} b =  -m \bar b  b + \frac{g^2(2 \kappa+1)}{ \kappa+1} (\bar b  b)^{ \kappa+1}, \ 
\eq
Now we will assume that in the lab frame $f(x,t)$ is independent of $t$ and of the form:
\bq \label{force2}
f_j(x) =  r_j e^{-iK_j x}, \quad j=1,2, 
\eq
with real parameters $r_j$ and $K_j$. 
In that case  from Eq. \eqref{fdens}, we have that $F^0 = 0$ and
\bq
\partial_{t} T^{00} + \partial_{x} T^{10}= 0, \label{c1}
\eq 
where
\begin{eqnarray} \label{de}
T^{00} &=& -\frac{i}{2} \left[  \bPsi  \gamma^1 \partial_x \Psi-\partial_x \bPsi \gamma^1 \Psi \right]+ m \bPsi \Psi - \frac{g^2}{ \kappa+1} (\bPsi \Psi)^{ \kappa+1} + \bPsi f+  \bar{f}  \Psi, \\  \label{t10}
 T^{10} &=& -\frac{i}{2} \left[\bPsi_{t}  \gamma^1 \Psi- \bPsi \gamma^1 \Psi_{t} \right]. 
\end{eqnarray}

Integrating Eq.\ (\ref{c1}), and under the assumption that $T^{10}(+\infty,t)-T^{10}(-\infty,t)=0$, then the energy of the driven NLD equation,  
\begin{eqnarray} \label{energy}
 E^{total}  &= &\int_{-\infty}^{+\infty}\, dx \,T^{00},  
\end{eqnarray}
is conserved.   

\section{Variational (collective coordinates)  Ansatz for the NLD equation with external driving forces} \label{sec5}

Our ansatz for the trial variational wave function is to assume that because of the smallness of the perturbation the main modification to our exact solutions to the NLD equation  (without driving forces)  is that the parameters describing the position $q(t)$, inverse width $\beta(t)$ and phase $\phi(t)$ become time dependent  functions.  We assume that the driving term is specified in the lab frame, and that the initial condition on the solitary wave is that it is a Lorentz boosted
exact solution moving with velocity $v$.  To describe the position of the solitary wave we introduce the parameter $q(t)$ which replaces $vt$ for the unforced case. We then let the width parameter $\beta$ and thus $\omega = \sqrt{m^2- \beta^2}$  become functions of time.  We next rewrite the phase of the exact solution as 
\bq
\omega t' = \gamma \omega( t-vx)  \rightarrow \phi(t) - p(t)(x-q(t))
\eq
to mimic our parametrization of the collective coordinates in the nonlinear Schr{\"o}dinger equation.  Next, we let 
$p(t) \equiv  \omega(t) \gamma(\dot q) \dot q  $  be determined from $\omega(t)$ and $q(t)$ and let the phase $\phi(t)$ be an independent collective variable.
That is, in Eq.\ (\ref{eq2.37}) we replace
\bq
vt \rightarrow q(t);~~ \beta \rightarrow \beta(t);  ~~~  \omega t'=  \gamma\omega( t - vx) \rightarrow \phi(t) -p(t)(x-q(t)) , 
\eq
where  $ p(t)= \gamma(t)   \omega(t) \dot q (t)$.   

Thus our trial wave function in component form is given by: 
\ba \label{eq4.2}
&&
\Psi_1(x,t) = \left( \cosh{\frac{\eta}{2}} A(z) 
+ i \sinh{\frac{\eta}{2}} B(z) \right) e^{-i \phi + i p(x-q)} , \nonumber \\
&&\Psi_2(x,t) = \left( \sinh{\frac{\eta}{2}} A(z) 
+ i \cosh{\frac{\eta}{2}} B(z) \right) e^{-i \phi + i p(x-q)} , 
\ea
where $z = \cosh \eta~( x-q(t))$. 
Note that $\omega$, which was a parameter in Eq. (\ref{eq2.33}), now is time dependent because of $\omega=\sqrt{m^2-\beta^2(t)}$.  
Using the trial wave function Eq.\ (\ref{eq4.2})  we can determine the effective Lagrangian for the variational parameters.
Writing the Lagrangian density as 
\bq
\mathcal{L} = \mathcal{L}_1 + \mathcal{L}_2+\mathcal{L}_3 , 
\eq
where 
\ba \mathcal{L}_1&&= \frac{i}{2} \left( \bPsi \gamma^\mu \partial_\mu \Psi - \partial_\mu \bPsi \gamma^\mu \Psi \right)  , \nonumber \\
{\cal L}_2 &&= - m \bPsi \Psi + \frac{g^2}{\kappa+1} (\bPsi \Psi)^{\kappa+1};~~ {\cal L}_3 = - \bPsi f -\barf  \Psi . 
\ea
Integrating over $x$ and changing integration variable to $z$ one obtains
\bq
L_1 = \int_{-\infty}^\infty  dx  \mathcal{L}_1 = Q \left( p {\dot q}+ {\dot \phi} - p \tanh \eta \right) - I_0  \cosh \eta - J_0 \tanh \eta , 
\eq
where the charge   
\bq
Q= \int dz[A^2(z)+B^2(z)],  
\eq
and the rest frame kinetic energy $I_0=H_1$
\bq
I_0 = \int dz \left( B' A-A'B  \right) = {H_1} , \label{izero}
\eq 
are given by Eqs. (\ref{A1}) and (\ref{A2}), respectively, in the Appendix.
 Here $ B'(x') = \frac{ dB(x')}{dx'}$, and  
\bq
J_0 = \int dz \left( {\dot  B} A- {\dot A}  B  \right)  ,  
\eq
and 
\bq {\dot  A} = \frac{ d A}{dt} = \frac{dA}{dz} \frac{dz}{dt}  , 
\eq
with a similar relation holding for $\dot B$.
Since $z=(x-q(t)) \cosh \eta$, we have 

\bq
\frac{dz}{dt}  = - {\dot q} \cosh \eta - z \tanh \eta { \dot \eta}  
\eq
and 

\bq
J_0 = - \cosh \eta  \dot{q} I_0 -{ \dot \eta}   \tanh{\eta}     \int dz 
` z  \left(  A  B' - B   A'   \right) . 
\eq 
The integrand in the second term is odd in $z$, so the integral vanishes and we are left with;
\begin{eqnarray} \label{L1}
L_1 &=& \int dx  \mathcal{L}_1 = Q \left( p {\dot q}+ {\dot \phi} - p \tanh \eta \right) - I_0 \left ( \cosh \eta -{\dot q} \sinh \eta \right), \\
L_2 &=& \int dx {\cal L}_2= -\frac{m}{\cosh \eta} I_1 
+ \frac{g^2}{(\kappa+1) \cosh \eta} I_2, \label{L2}
\end{eqnarray}
where 
\ba
I_1&& = \int dz \left( A^2(z)-B^2(z)\right)= \frac{H_2}{m}; \qquad I_2=\int dz \left( A^2(z)-B^2(z) \right)^{\kappa+1}=\frac{\kappa+1}{g^2} H_3=\frac{\kappa+1}{g^2 \kappa} H_1.  \label{i2} 
\ea
For $L_3$ we have
\ba
L_3 &&= -  2 \int dx Re \left[f^\star_1 \Psi_1 - f_2^\star \Psi_2 \right]= \frac{1}{\gamma} \int dz {\mathcal L}_3. \nonumber \\
\ea

In what follows we make the simplification $\nu_j=0$ in Eq. (3.7) and
obtain for the integrand
\begin{eqnarray}
{\mathcal L}_3 &=& - 2 
r_1 \cos(\phi-K_1 q) \left\{ \cosh \frac{\eta}{2}  A(z) \cos \frac{p+K_1}{\gamma} z
-  \sinh \frac{\eta}{2}  B(z) \sin \frac{p+K_1}{\gamma} z \right\}
 \nonumber \\ 
&+& 2 r_2 \cos(\phi-K_2 q) \left\{ \sinh \frac{\eta}{2}  A(z) \cos \frac{p+K_2}{\gamma} z
-  \cosh \frac{\eta}{2}  B(z) \sin \frac{p+K_2}{\gamma} z \right\} , 
\end{eqnarray}
where  we have not included terms that are  odd in $z$.
Performing the integration we get
\begin{eqnarray} \label{4.15}
L_3 &=& - 2 
\frac{r_1}{\gamma} \cos(\phi-K_1 q) \left( \cosh \frac{\eta}{2} J_1
-  \sinh \frac{\eta}{2}  N_1 \right) 
+ 2 \frac{r_2}{\gamma} \cos(\phi-K_2 q)  \left( \sinh \frac{\eta}{2}  J_2
-  \cosh \frac{\eta}{2}  N_2 \right), \label{4.15a} \\
J_j(\omega,\dot{q}) &=& \int dz A(z)  \cos \frac{p+K_j}{\gamma} z = 
\frac{\pi \cos b_j}{g \sqrt{\omega} \cosh a_j \pi}, \quad a_j=\frac{p+K_j}{2 \beta \gamma}, 
\quad b_j= a_j \cosh^{-1} m/\omega, \label{4.15b} \\
N_j(\omega,\dot{q}) &=& \int dz B(z)  \sin \frac{p+K_j}{\gamma} z = 
\frac{\pi \sin b_j}{g \sqrt{\omega} \cosh a_j \pi}.
\end{eqnarray}
The integrals $I_1$, $I_2$, $J_j$ and $N_j$ are done exactly in the 
Appendix. 
Putting all terms together and using the fact that $ \dot q = v = \tanh \eta$ we obtain:
\ba \label{l3}
L&& = Q {\dot \phi}  - I_0 ~ \sech \eta 
  -{m} I_1  \, {\sech \eta} +\frac{g^2}{\kappa+1}  I_2 \, \sech \eta + L_3, \nonumber \\
 L_3 && = - 
  \frac{2\pi}{g \gamma \sqrt{\omega}}  \left\{
  \frac{r_1 \cos(\phi-K_1 q)}{ \cosh a_1 \pi} 
 C_1  - \frac{r_2 \cos(\phi-K_2 q)}{ \cosh a_2 \pi} 
 S_2 
  \right\}, \nonumber \\ \label{eq4.16}
C_j && = \cosh \frac{\eta}{2} \cos b_j 
  -\sinh \frac{\eta}{2} \sin b_j, \quad   
S_j  = \sinh \frac{\eta}{2} \cos b_j 
 -\cosh \frac{\eta}{2} \sin b_j, \quad j=1,2.
\ea
Since we are using the exact solutions  of the NLD equation as our trial wave functions for the forced problem, the integrals
$I_0, I_1$ and $I_2$ are related since
 for the NLD equation without the presence of external forces, the solitary wave  with $v=0$ obeys the relationship 
\cite{LeeKuoGavrielides1975}  
\bq
\omega \psi^\dag \psi - m \bpsi \psi + \frac{g^2}{\kappa+1}
(\bPsi \Psi)^{\kappa+1} =0.
\eq
For our problem this converts into 
\bq
\omega(A^2+B^2) -m(A^2-B^2) + \frac{g^2}{\kappa+1} (A^2-B^2)^{\kappa+1} =0.
\eq
Integrating this relationship we obtain:
\bq
mI_1 - \frac{g^2}{\kappa+1} I_2 - \omega Q=H_2 -\frac{H_1}{\kappa}-\omega Q=0.  \label{relation1}
\eq
Using this relation to replace $I_1$ and $I_2$ in $L$ we have  
\bq
L = Q {\dot \phi}  - \frac{1}{\gamma}  (I_0 + \omega Q)  - { U }(q ,\dot q, \beta,\phi) = Q {\dot \phi}  - \frac{M_0}{\gamma}   - { U }(q ,\dot q, \beta, \phi)  ,~~ \label{Lag}
\eq
where $U = - L_3$, and $M_0=I_{0}+\omega Q$ is the rest frame energy of the solitary wave for $\kappa=1$. 
 
From Eq. \eqref{df5} we can calculate the dissipation function $F$ for the CC equations.
We find
\ba
F&& = 2 \mu \int_{-\infty}^{+\infty} dx Im (\Psi^\dag \partial_t \Psi) \nonumber \\
&&= 2 \mu  \int_{-\infty}^{+\infty} \frac {dz}{\cosh \eta} \left[ \sinh \eta \,(A \dot B - B \dot A) - \cosh \eta \,(p \dot q + \dot \phi) (A^2+B^2) \right]. 
\ea
We recognize the integrals as being related to $J_0 = - \cosh \eta  \dot q I_0$ and $Q$, so we obtain
\bq
F = -2 \mu  \left[ I_0 \sinh \eta {\dot q} + Q (p \dot q + \dot \phi) \right].
\eq

We can simplify this by introducing the boosted rest frame mass: 
\bq \label{eqM}
M =   \gamma M_0 \equiv \gamma (I_0 + \omega Q) 
\eq
and use the definition of $p(t) = \gamma \omega \dot q$ so that 
\bq
F=- 2 \mu (M {\dot q}^2 + Q \dot \phi).
\eq

This is the relativistic generalization of our expression that we found for the forced NLS equation \cite{MertensQuinteroBishop2010}. 
Now we are ready to derive Lagrange's equations for the collective coordinates using Eq. (\ref{Lag}).
From
\bq \label{eq4.24}
\frac{d}{dt} \frac {\partial L}{\partial \dot q} -  \frac{\partial L} {\partial q} = \frac {\partial { F}}{\partial \dot q}, 
\eq
we obtain
\bq \label{eq4.25}
\frac{d}{dt} \left( M \dot q \right)  = {F} _{eff}, \eq
where
\ba \label{eq4.25a}
 { F} _{eff}& =& \frac{d}{dt} \frac {\partial {U} }{\partial \dot q} -  \frac{\partial {U} } {\partial q}+ \frac {\partial  F}{\partial \dot q} .  
\end{eqnarray}
We also have a contribution from dissipation from the equation
\bq
\frac{d}{dt} \frac {\partial L}{\partial \dot \phi} -  \frac{\partial L} {\partial \phi} = \frac {\partial {F}}{\partial \dot \phi},
\eq 
which gives us a first-order differential equation for $\omega$
\bq \label{omegadot}
\dot Q = Q'(\omega) \dot \omega = - 2 \mu  Q-  \frac{\partial U}{\partial \phi}, 
\eq
where the prime denotes the derivative with respect to $\omega$.

As $L$ does not depend on $\dot{\beta}$, the final Lagrange equation is 
$\partial L/\partial \beta=0$. After changing to the variable $\omega=\sqrt{m^2-\beta^2}$ we have  
\bq
\frac{\partial L} {\partial \omega} = 0. \label{final} 
\eq

This leads to  a first-order differential equation for $\phi$

\bq \label{relation3}
Q'(\omega) \dot \phi = \frac{1}{\gamma}  M_0'(\omega) +\frac{\partial U}{\partial \omega} . 
\eq
Here $U= - L_3$, and $L_3$ is given by Eq. \eqref{l3}. The second-order ODE Eq. \eqref{eq4.25} and the first-order ODEs Eqs. \eqref{omegadot} and \eqref{relation3} will be solved numerically in Sec. VI and compared with our simulations.

\section{Spectrum of the linear excitations (phonons)} \label{secphonons}
Similar to the case of the forced NLS equation \cite{MertensQuinteroBishop2013} the spinor force Eq. (3.21) excites not only soliton excitations, but also plane wave phonons. (We will use the word phonons for the linear excitations).

The general solution of the linearized NLD, Eq. (3.6) without damping ($\mu=0$), reads
\bq
\Psi_{ph} = a e^{i(kx-\omega_{ph} t)}+ b e^{-i K_1x} + c e^{-i K_2 x},
\eq
with arbitrary, but small $a$, and the phonon dispersion curve
\bq
\omega_{ph} = \sqrt{k^2 + m^2}.
\eq
The first term in Eq. (5.1)  is the solution of the homogeneous equation, while the second and third terms represent a particular solution with $b$ and $c$ corresponding to the spinors:
\bq
 b= \frac{r_1}{\Omega_1^2} \left( \begin{array}{cc}
    -m \\ 
    K_1
  \end{array}
    \right ) ~;~~
    c= \frac{r_2}{\Omega_2^2} \left( \begin{array}{cc}
    K_2 \\ 
    m
  \end{array}
    \right ) ~~ 
    \eq
    and the frequencies
 
\bq
\Omega_1 = \sqrt{K_1^2+m^2}, ~~\Omega_2 = \sqrt{K_2^2+m^2} \, . 
\eq
These predicted frequencies are clearly identified in the spectrum of the charge $Q(t)$, which is obtained in our simulations (Sec. VI). 

The phonon modes are also seen indirectly in the spectrum of the maximum of the charge density $\rho(x,t)$.  This is a {\em local} quantity which is used for the computation of the soliton position $q(t)$, in contrast to the {\em global} quantity $Q(t)$ which is obtained by integration over the whole system.  The phonon frequencies are observed in the differences $\Omega_{1,2}- \omega_s$, where $\omega_s$ is the frequency of the intrinsic soliton oscillation found in Sec. VI.  Note that $\omega_s$ is also observed in the discrete Fourier transform (DFT) of $Q(t)$, together with $\Omega_1$ and $\Omega_2$.

\section{Comparison of Collective Coordinate results with simulations} \label{sec6}

We have the four parameters $r_1,r_2,K_1,K_2$ in the forces, and four initial conditions (ICs) $q(0)=q_0,~ \dot q(0)=v_0,~ \omega(0) = \omega_0$ and $ \phi(0) = \phi_0$ for the ODEs Eqs. (4.31), (4.34) and (4.26).  In our simulations we use the exact moving solitary wave solution Eq. (2.8) of the unperturbed NLD equation, with the replacements in Eqs. (4.2) and the same ICs as for the ODEs. 
	As the space of the parameters and ICs is 8-dimensional, we must find out which regions in this space are relevant for us.  For this reason we impose the following restrictions: 
\begin{enumerate}
\item The forces must be sufficiently small. This concerns the amplitudes $r_1$ and $r_2$ of the components of the spinor force. The simulations reveal that when $r_i \ge 0.03$ a background appears on both sides of the soliton.  However, when $r_1 =0.01$ and $r_2 < r_1$ no background appears (see Fig.1).  This characteristic is maintained even for very long integration times $t_f = 3000$. 
\item $ |r_2|$ should be smaller than $|r_1|$ because the second component of a spinor is the so-called ``small component''.  As we have chosen $r_1=0.01$ to satisfy our first restriction, we will use in our simulations $r_2 = \pm 0.001, \pm 0.005, \pm 0.009$.
\item $K_1$ and $K_2$ must be sufficiently small.  The length scales $l_1 = 2 \pi/ |K_1|$ and $l_2= 2 \pi/ |K_2|$ on which the force components vary, must be much larger than the soliton width $b$. Otherwise the soliton will not behave like a particle.  We choose $K_1 =0.5$, i.e. $l_1 = 12 \ll b \approx 5$ (when $\omega_0 = 0.9)$, and 
$K_2 = - 0.1$, i.e. $l_2 = 60 \ll b$. Larger values, for example $K_1 = 1.0, K_2 = -0.1$, yield qualitatively different results, since $l_1 = 6.2$ is of the same order as
$b=5$.
\item $K_1$ and $K_2$ should have opposite signs.  If this is not the case (e.g. $K_1=0.6$, $K_2 = 0.1$) and if $v_0=0$, then there are very slow oscillations of the soliton position whose period $T_q$ is larger than the maximum integration time which is $t_f = 3000$, so these cannot be captured in the numerical analysis.
For the choice $K_1 = 0.5$, $K_2 = -0.1$ the period of this slow oscillation is about the same as $t_f$, so it can be captured by our numerical study.
\item $ |K_1|$ and $|K_2|$ must differ strongly in order to see the phonon peaks in the numerical study.  The phonon peaks can be seen in the spectrum of the charge $Q(t)$ which is a global quantity (see Sec.~V). When $K_1 =K_2$ there is only one peak (see \cite{us1}). However for $K_1 \neq K_2$  two peaks are expected (here we choose $m=1$) at $\Omega_1 = \sqrt{m^2+K_1^2}$ and  $\Omega_2 = \sqrt{m^2+K_2^2}$.  In order to have well separated peaks. $|K_1|$ and $|K_2|$ must differ sufficiently.  We take for example $K_1 = 0.5$ and $K_2 = -0.1$ which yields $\Omega_1= 1.1180$ and $\Omega_2 = 1.0049$.  The frequency difference is $\Delta \Omega = 0.1131$ which is visible. Note that we cannot take a  larger value for $K_1$ because of point 3. 
\item The initial velocity $v_0$ must be small or zero.  If $v_0$ is not much smaller than one, the soliton soon reaches one of the boundaries of our numerical simulation.
Therefore we choose $v_0=0.1$.  For this choice the soliton covers 80 space units in the integration time $t_f=800$ and does not yet reach the boundary at $x=100$.
For our simulations we choose the system to be in the interval $\left[ -100, 100 \right]$. 
	For $v=0$, the soliton travels only a short distance, therefore a final integration time of $t_f = 3000$ can be taken, which is technically the maximum time in our simulation program.  The computational cost taking the maximum time is huge because our fourth-order operator splitting method that we have used earlier \cite{ShaoQuinteroMertens2014} \cite{us1}  requires that the spatial spacing $h= \frac {\tau}{12}$, where we choose the time step  $\tau= 0.025$. This implies that the number of grid points is $96,000$ for the above system size. 
	
\end{enumerate}
 
As parameters of the NLD equation (3.61) we choose $m=1,g=1,\kappa=1$ and $\mu=0$. For the spinor force Eq. (3.7) we take into account the above six points. and choose $r_1=0.01, r_2=\pm 0.005, K_1=0.5$ and $K_2=- 0.1$. Moreover, we restrict ourselves to time independent forces and set $\nu_1=\nu_2=0.$ As ICs we first take $q(0)=q_0=0, \dot q(0) = v_0 =0, \omega(0)=\omega_0=0.9$, and $\phi(0)= \phi_0 = \pi/2$.  Other ICs will be considered below.
Figures 2(a), (b) show simulation results for the charge $Q(t)$ and its Discrete Fourier Transform (DFT).  The highest peak is situated at $\omega_s=0.9006$. This can be identified as the frequency of the intrinsic soliton oscillations, because our CC theory yields $\omega_s^{cc} = 0.8985$ for the oscillations of all collective variables.

The two peaks at $\Omega_1=1.1268$ and $\Omega_2=1.0032$ agree well with the predicted phonon peaks at $\Omega_1=1.1180$ and $\Omega_2=1.0049$, see Eq. (5.4) and  the above point 5. The fourth peak at $0.0020944$ is identical to the smallest frequency $2 \pi/t_f$ that appears in the DFT, where  $t_f$=3000 is the integration time. 

Next we discuss the translational motion of the soliton.  Our CC theory predicts two scenarios: 
\begin{enumerate}
\item The soliton is trapped and oscillates very slowly around a mean value. In Fig. 3(a) the oscillation amplitude $a_q^{cc}$ is about $0.7$, the period is $T_q^{cc} \approx 1000$ and the mean value of the position is equal to the initial value $q_0=0.$ Comparing with the simulation results in Figs. 2(c), (d), we see only a qualitative agreement.  However, the rapid oscillations, which are superimposed on the slow ones, agree quite well: $\omega_s^{cc} = 0.8985$ compared to $\omega_s=0.9006$ from the DFT of $q(t)$ shown in Fig. 2(d).  The amplitudes of these oscillations are about $0.14.$ There is another peak in the spectrum of $q(t)$ at $0.2262$, which is exactly equal to the difference between the upper phonon peak at $\Omega_1=1.1268$ and the soliton peak at $\omega_s=0.9006$. Via this difference the phonon frequency is observed {\it indirectly}. This is because the soliton position $q(t)$ is a {\it local} quantity, in contrast to the {\it global} quantity $Q(t)$ in which the phonons are observed {\it directly}, see above.  The lower phonon peak at $\Omega_2 = 1.0032$ is weaker (see Fig. 2(b)) and therefore it is not visible in the above difference. Fig. 2(e) exhibits the maximum of the charge density $\rho(x,t)$ as a function of time.  This is what we call  the amplitude of the soliton in the CC language which is $a=2 (m-\omega(t))/g^2$ and is a local quantity.  Consequently its spectrum in Fig. 2(f) also has a peak at the difference between the phonon and the soliton peak. The amplitude of the oscillations in Fig. 2(e) is roughly $0.012$ which agrees  rather well with the CC result $0.0092$. 
\item The second scenario which the CC theory predicts for the translational motion of the soliton consists of the following: the soliton performs oscillations
around a mean path given by $\barv_{cc} t$, in Fig. 3(c),   $\barv_{cc} = -0.0131.$ In Fig. 3(d), in order to better see these oscillations, we plot $q(t) - \barv_{cc} t $. 
The oscillations consist of very slow ones with frequency $\omega_q^{cc} \approx 0.00785$ and amplitude $a_q^{cc} = 1.07$,  and rapid ones with the intrinic soliton frequency $\omega_s^{cc} = 0.9048$ and amplitude $a_s^{cc} \approx 0.2$ (Fig. 3(d)).  When compared with the simulation results there is only a qualitative agreement concerning the very slow oscillations (Figs. 4(c), (d)).  However the rapid oscillations again agree well: $\omega_s = 0.9027$ and $a_s = 0.15$.  Moreover, the frequency and amplitude of the soliton  amplitude oscillations as defined by the charge density (Figs. 4(e), (f)) agree well with the CC results.
\end{enumerate}
Because of the space dependent spinor forces 
$$ f_j= r_j \exp (-i K_j x); ~~ j=1,2$$
the system is not homogeneous. 
Therefore the time evolution of the coordinates depends on the initial soliton position $q_0$.  In Table I we show how the characteristics of both the translational and the intrinsic dynamics of the soliton depend on $q_0$ which is given in units of $l_1 = 2 \pi/K_1$.  For broad intervals of $q_0$ the soliton travels in one direction.  These intervals alternate with other broad intervals where the soliton travels in the opposite direction.  However, in between there are narrow intervals in which the soliton is trapped.  Here both the period $T_q^{cc}$ and the amplitude $a_q^{cc}$ of the oscillations are  considerably larger than they are in the travel intervals (Table I).

This pattern of alternating intervals depends on the initial phase $\phi_0$, but is always very similar. As to the IC $\omega_0$, we restrict ourselves to the non-relativistic regime and take $\omega_0=0.9$ which is close to $m=1$. Here we expect stable solitons because in the non-relativistic limit we approach the Nonlinear Schr\"{o}dinger (NLS) 
Equation.  In a future work we plan to consider the fully relativistic regime (e.g., $\omega_0 = 0.5$) and the ultrarelativistic regime (e.g., $\omega_0=0.1$). 
\begin{table}[htdp]
\caption{Variation of $q_0$ for ICs $v_0=0, \phi_0=0, \omega_0=0.9$.  Parameters: $r_1=0.01, r_2=0.00 5, K_1=0.5, K_2=-0.1, t_f=3000$.}
\centering
\begin{tabular}{|c|c|c|c|c|c|}
\hline
$q_0/l_1$ & type of motion & $ {\bar v}_{cc}$ & $T_q^{cc}$ & $a_q^{cc}$ & $\omega_s^{cc}$ \\
\hline \hline
0 & travel & +0.06148 & 167 & 0.10 & 0.8650 \\
\hline
0.05 & travel & +0.05843 & 176 & 0.12 & 0.8650 \\
\hline
0.1 & travel & +0.04953 & 214 & 0.14 & 0.8713 \\
\hline
0.2 &  travel &+0.01503  &750  & 0.85  & 0.8922 \\
\hline
0.25 & trapped &- & 1100 & 3.65 & 0.8943 \\
\hline
0.3 & travel  &-0,01579  & 600  & 0.75  &  0.9090 \\
\hline
0.4 & travel   & -0.05169  &200  & 0.155 & 0.9278  \\
\hline
0.5 &  travel & -0.06514 & 158  & 0.135 &  0.9341 \\
\hline
0.6 & travel  & -0.05462  & 188  & 0.14  & 0.9278  \\
\hline
0.7  & travel &-0.02601  & 428  & 0.4  & 0.9131  \\
\hline
0.75  & trapped  & - & 1800 &4.7  &  0.9027 \\
\hline
0.8 &  travel  &+0.02036  & 500 & 0.50 & 0.8880  \\
\hline
0.9 & travel   & +0.04835  & 214  & 0.135  & 0.8712  \\
\hline
1.0 & travel & +0.05886 & 176  & 0.12 & 0.8650 \\
\hline \hline
\end{tabular}
\label{default}
\end{table}

\begin{figure}[ht!]
\begin{center}
\begin{tabular}{cc}
\ & \\
\includegraphics[width=8.0cm]{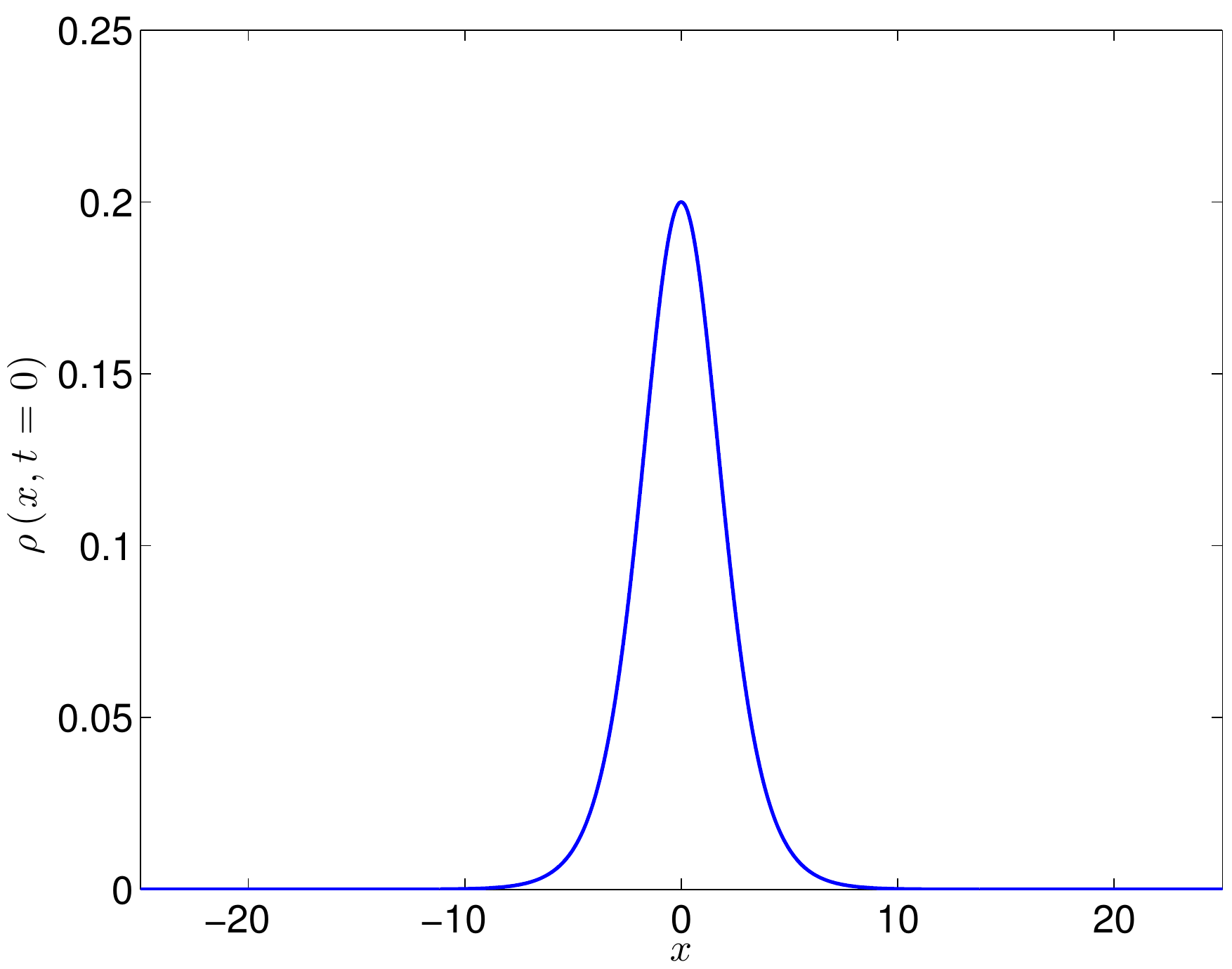}  & 
\quad \includegraphics[width=8.0cm]{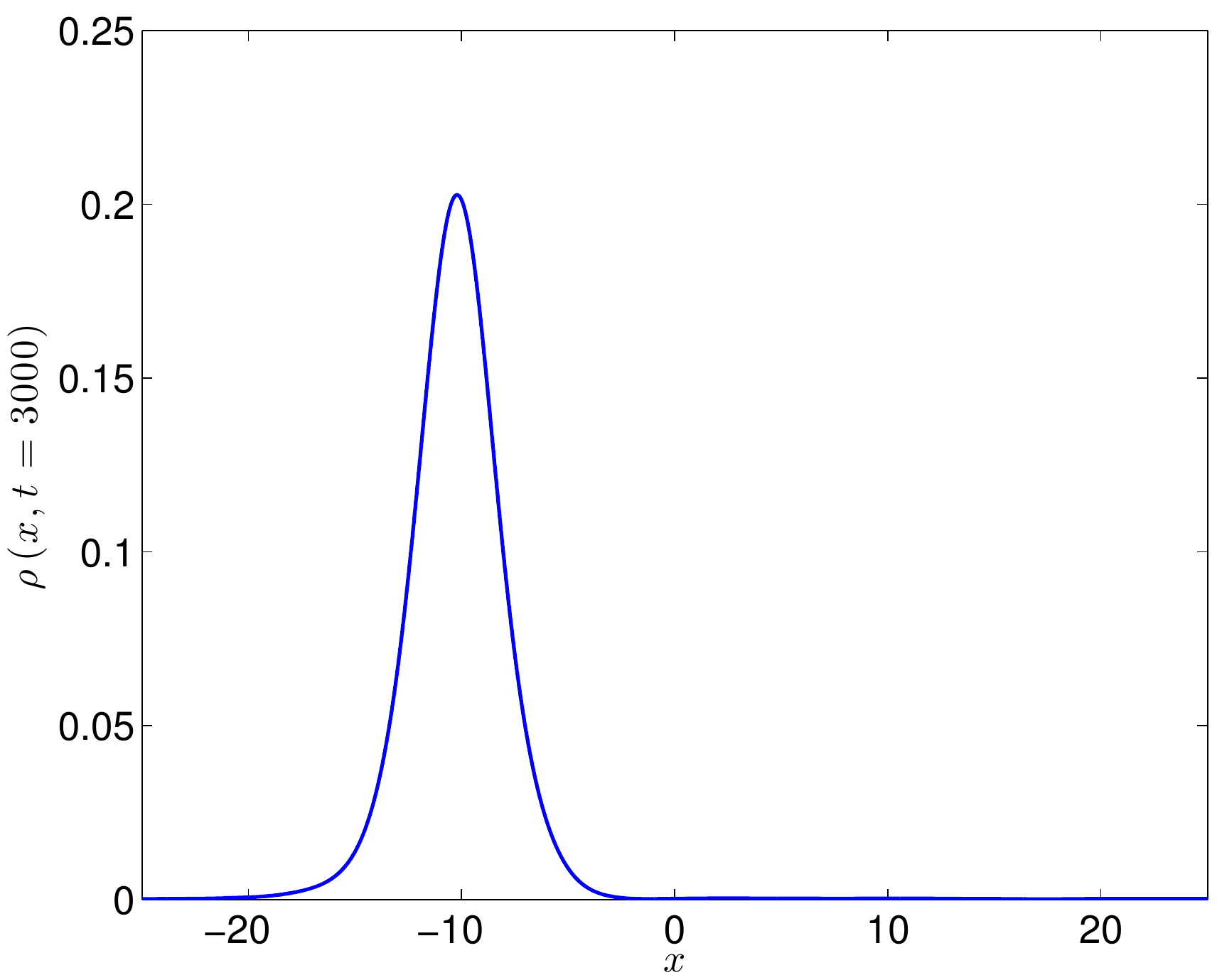} \\
\end{tabular}
\end{center}
\caption{Snapshots of the soliton profile at different times.
Left  panel: charge density $\rho(x,t=0)$. 
Right panel: charge density $\rho(x,t=3000)$. 
Parameters: $r_1=0.01, r_2=0.005, K_1 =0.5, K_2=-0.1.$  Initial conditions: $q_0=0,v_0=0, \phi_0=\pi/2, \omega_0=0.9$. }
\label{fig1} 
\end{figure}

\begin{figure}[ht!]
\begin{center}
\begin{tabular}{cc}
\ & \\
\includegraphics[width=7.0cm]{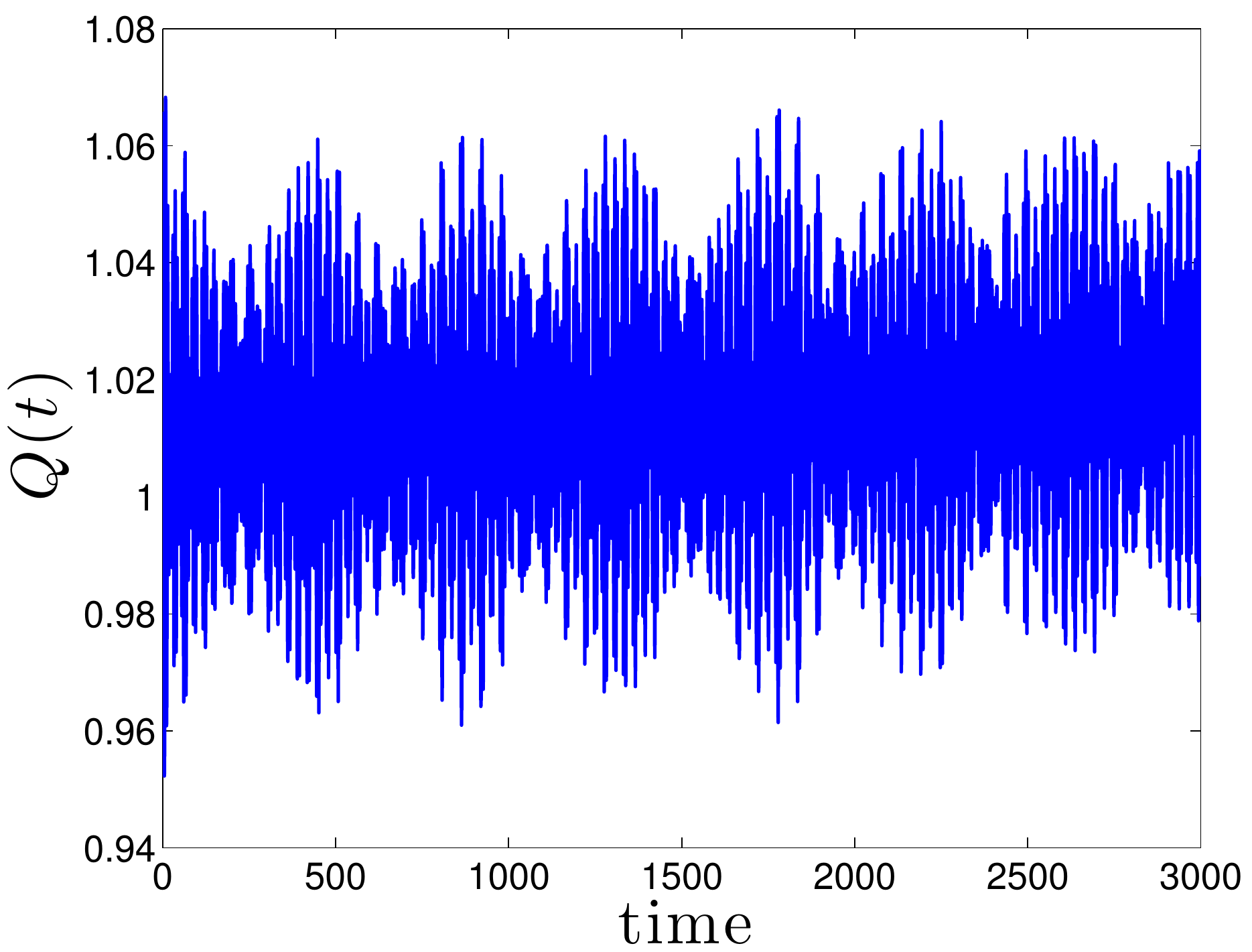}  & 
\quad \includegraphics[width=7.0cm]{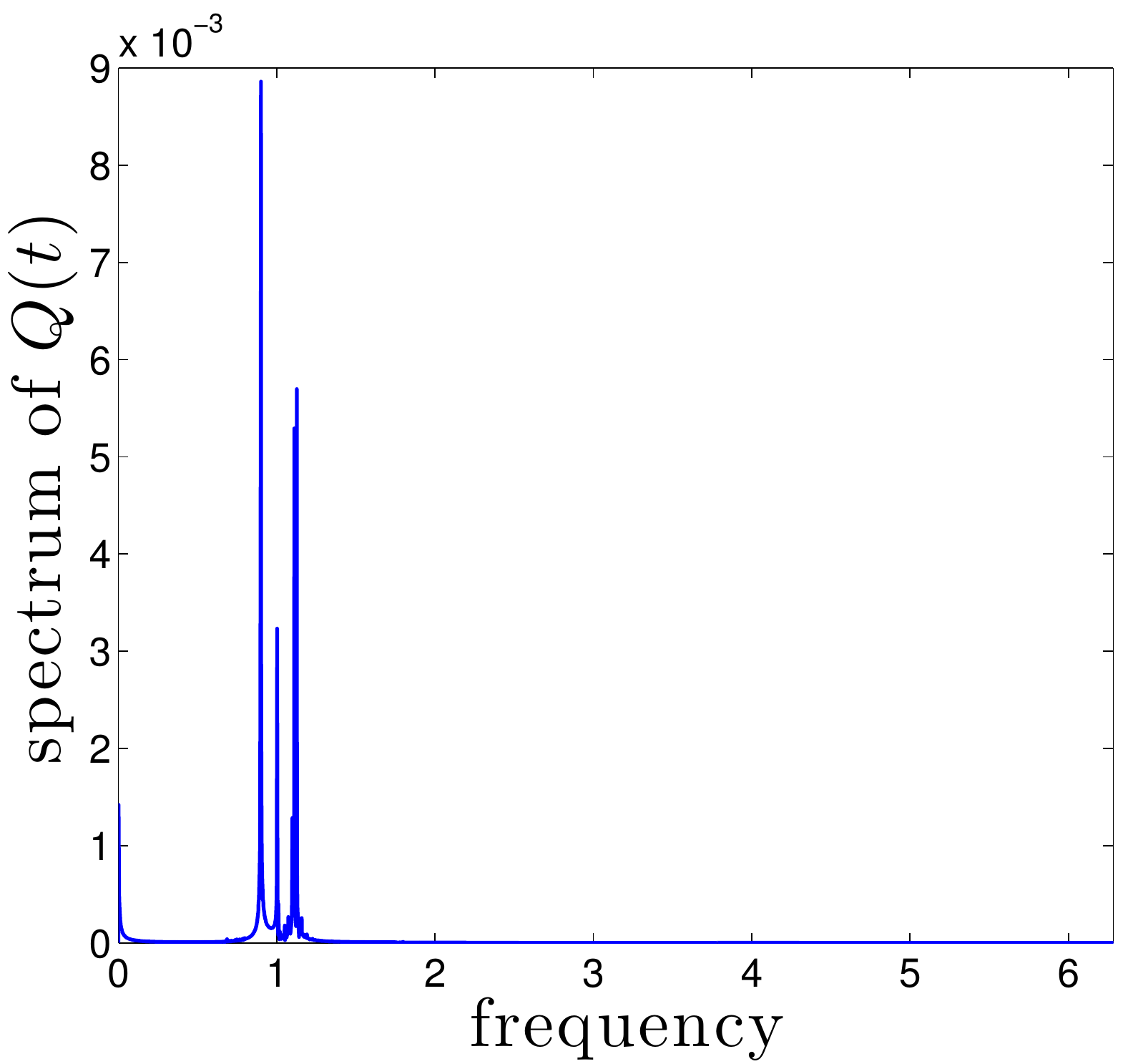} \\
 \\
\includegraphics[width=7.0cm]{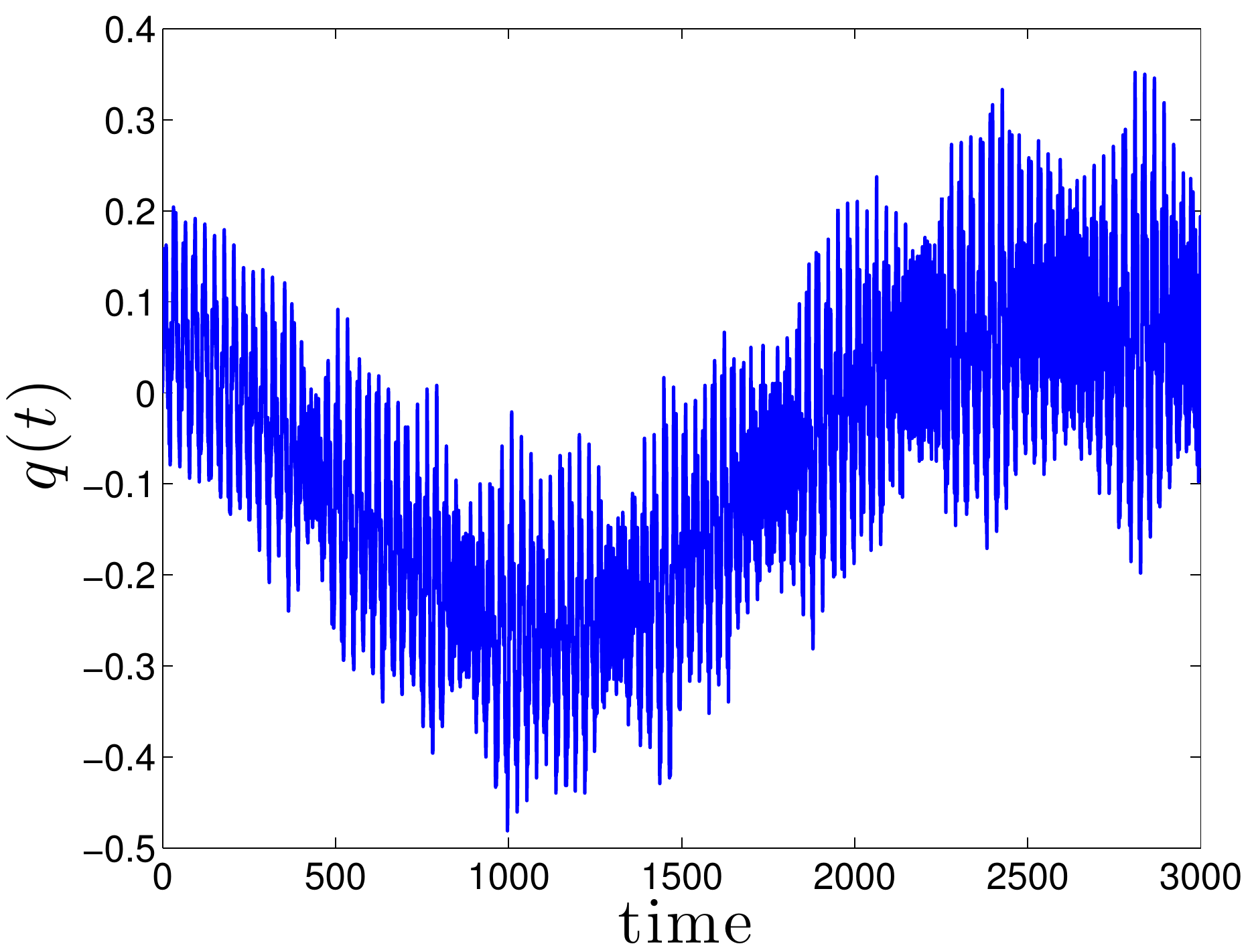}  & 
\quad \includegraphics[width=7.0cm]{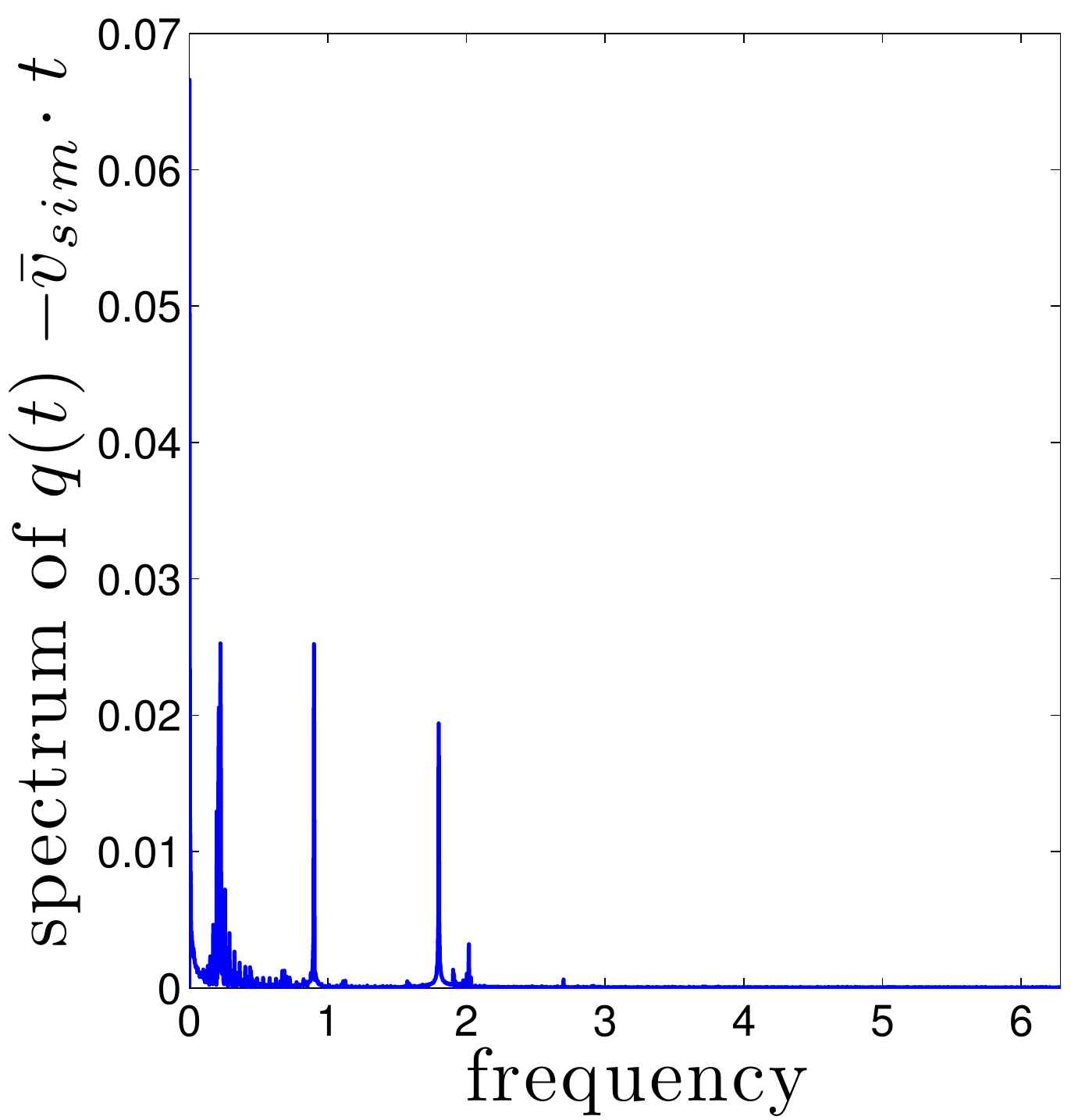}
\\ 
\includegraphics[width=7.0cm]{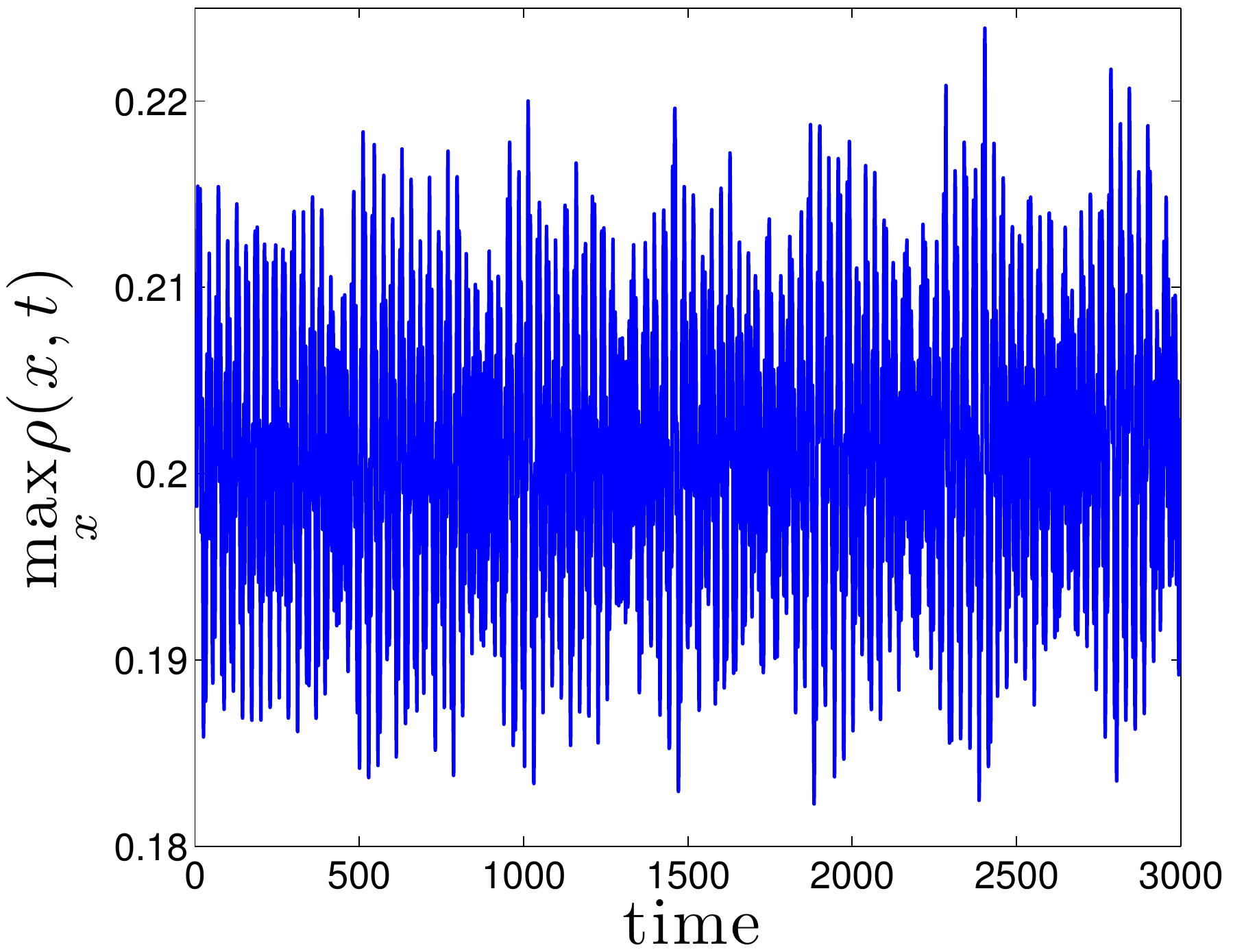}  & 
\quad \includegraphics[width=7.0cm]{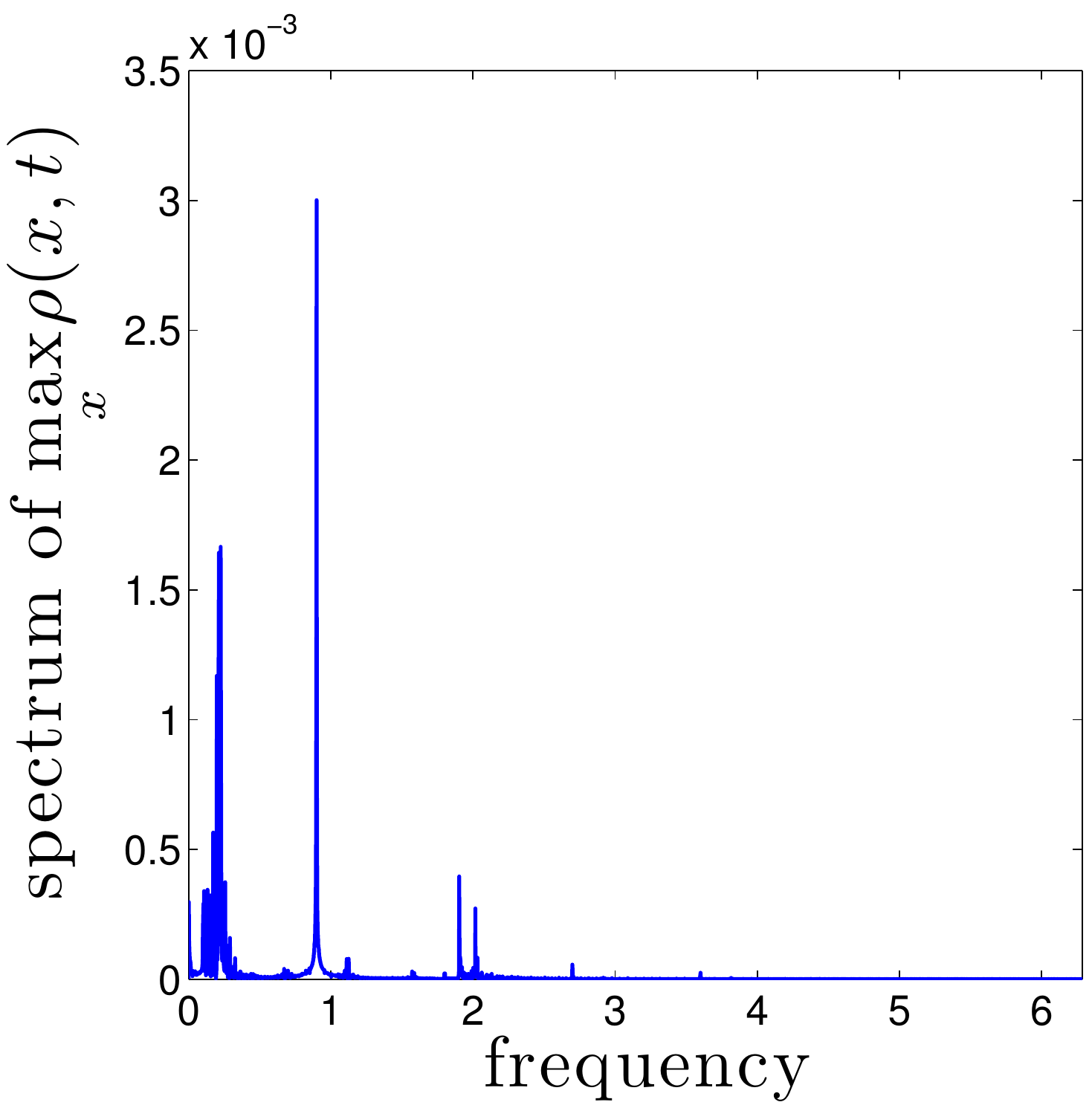}
\\
\end{tabular}
\end{center}
\caption{Simulation results: Oscillations of a trapped soliton. Same parameters and initial conditions as in Fig.~1 except $r_2=-0.005$.  
Left upper panel: Charge  $Q(t)$. Right upper panel: spectrum of $Q(t)$ with  peaks  at the frequencies $ 0.9006, 1.1268,1.0032$ and $0.0020944.$  Left middle panel: position $q(t)$. Right middle panel: spectrum of $q(t)$  with peaks at the frequencies $0.0020944, 0.2262, 0.9006 $ and $1.7991$.   
Left lower panel:  $max_x \rho(x,t)$.  Right lower panel:  spectrum of $max_x  \rho(x,t)$ with peaks  at $0 .9006 $ and $ 0.2262.$}
\label{fig2} 
\end{figure}

\begin{figure}[ht!]
\begin{center}
\begin{tabular}{cc}
\ & \\
\includegraphics[width=7.0cm]{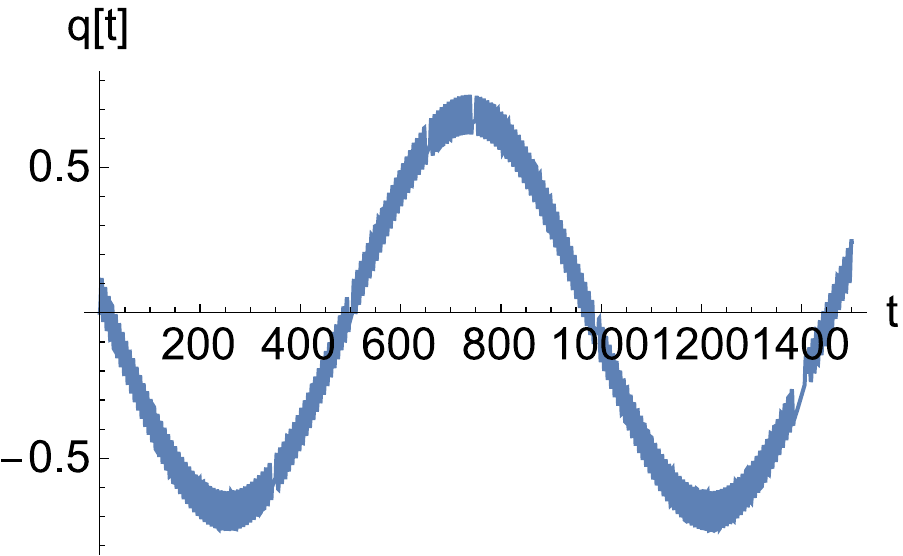}  & 
\quad \includegraphics[width=7.0cm]{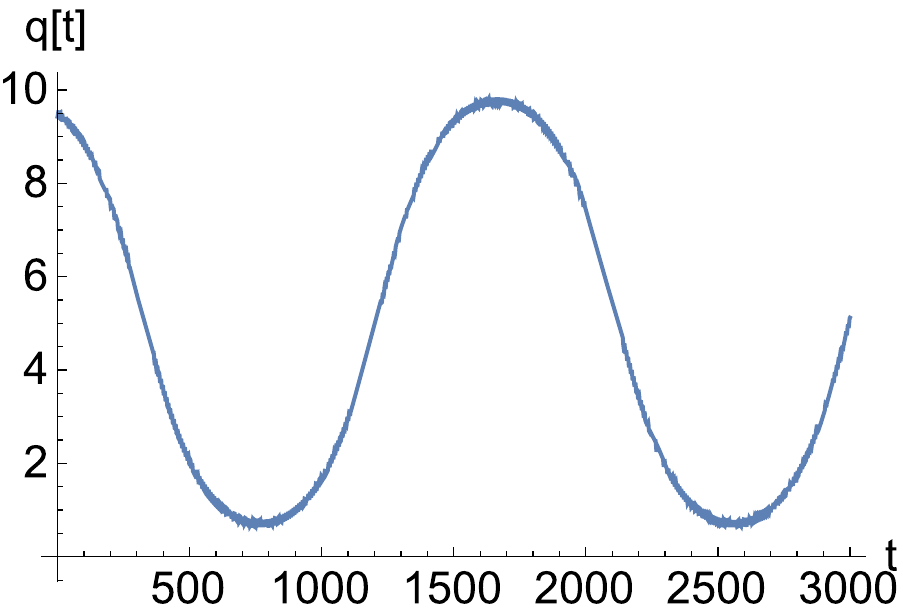} \\
 \\
\includegraphics[width=7.0cm]{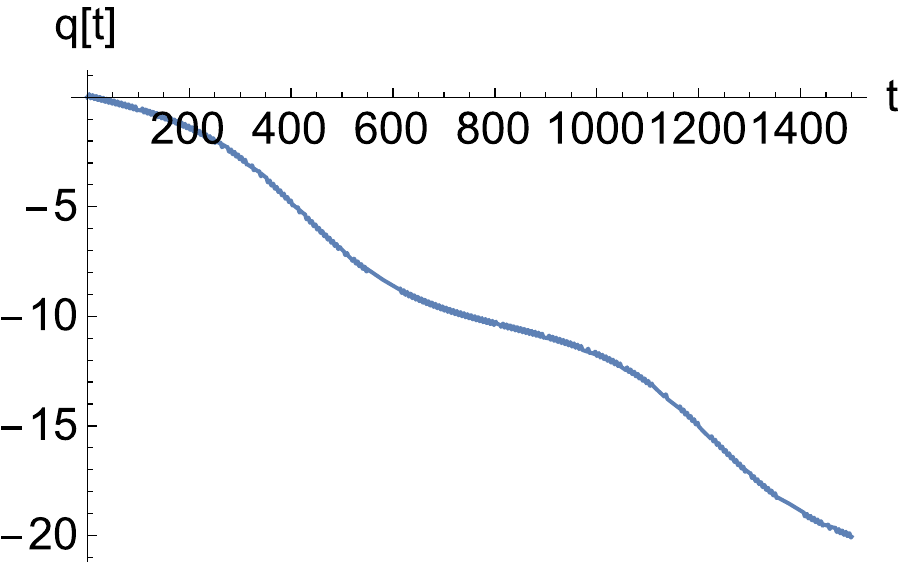}  & 
\quad \includegraphics[width=7.0cm]{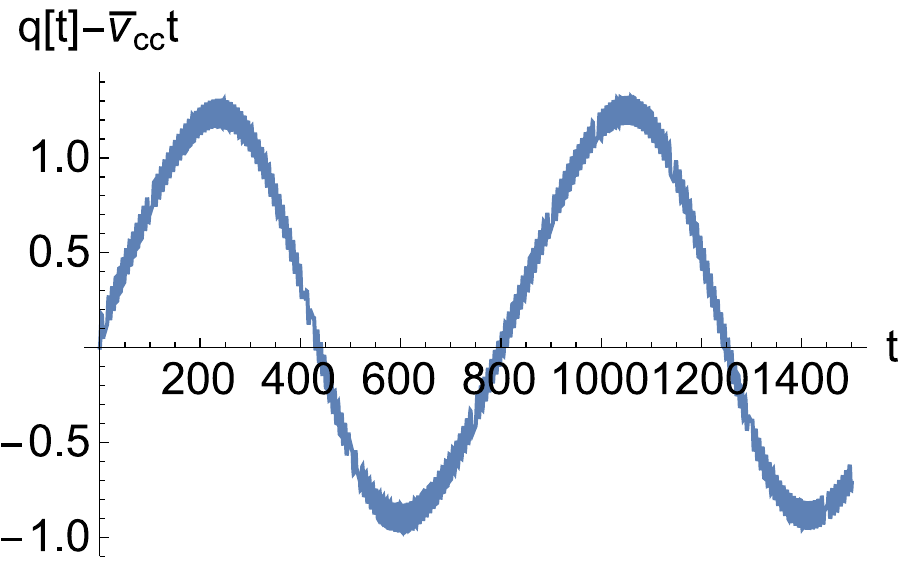}
\end{tabular}
\end{center}
\caption{Results of Collective Coordinates Theory: Oscillations of a trapped or a traveling  soliton.  Panel (a): Same parameters and initial conditions as in Fig.~1 except $r_2=-0.005$.  Panel (b):   Same parameters and initial conditions as in Fig.~1 except $q_0= 0.75 l_1$ and $\phi_0=0$.  Panels (c), (d): Same parameters and ICs as in Fig.~1.}
\label{fig3} 
\end{figure}

Figure \ref{fig3}  shows the  results of the collective coordinate theory for three initial conditions. Figures 3(c) and 3(d) pertain to a traveling wave case discussed in Fig.1. 

\begin{figure}[ht!]
\begin{center}
\begin{tabular}{cc}
\ & \\
\includegraphics[width=7.0cm]{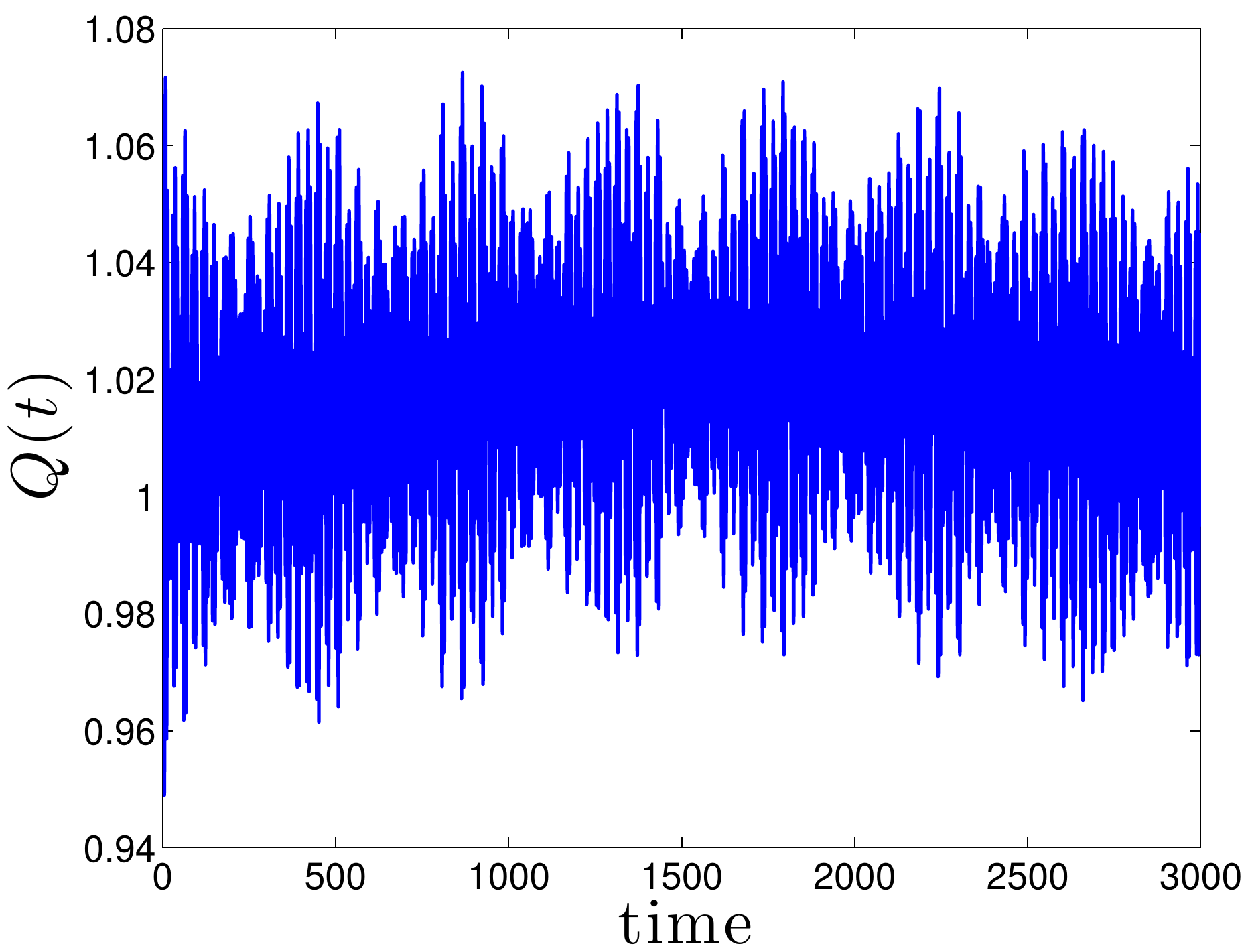}  & 
\quad \includegraphics[width=7.0cm]{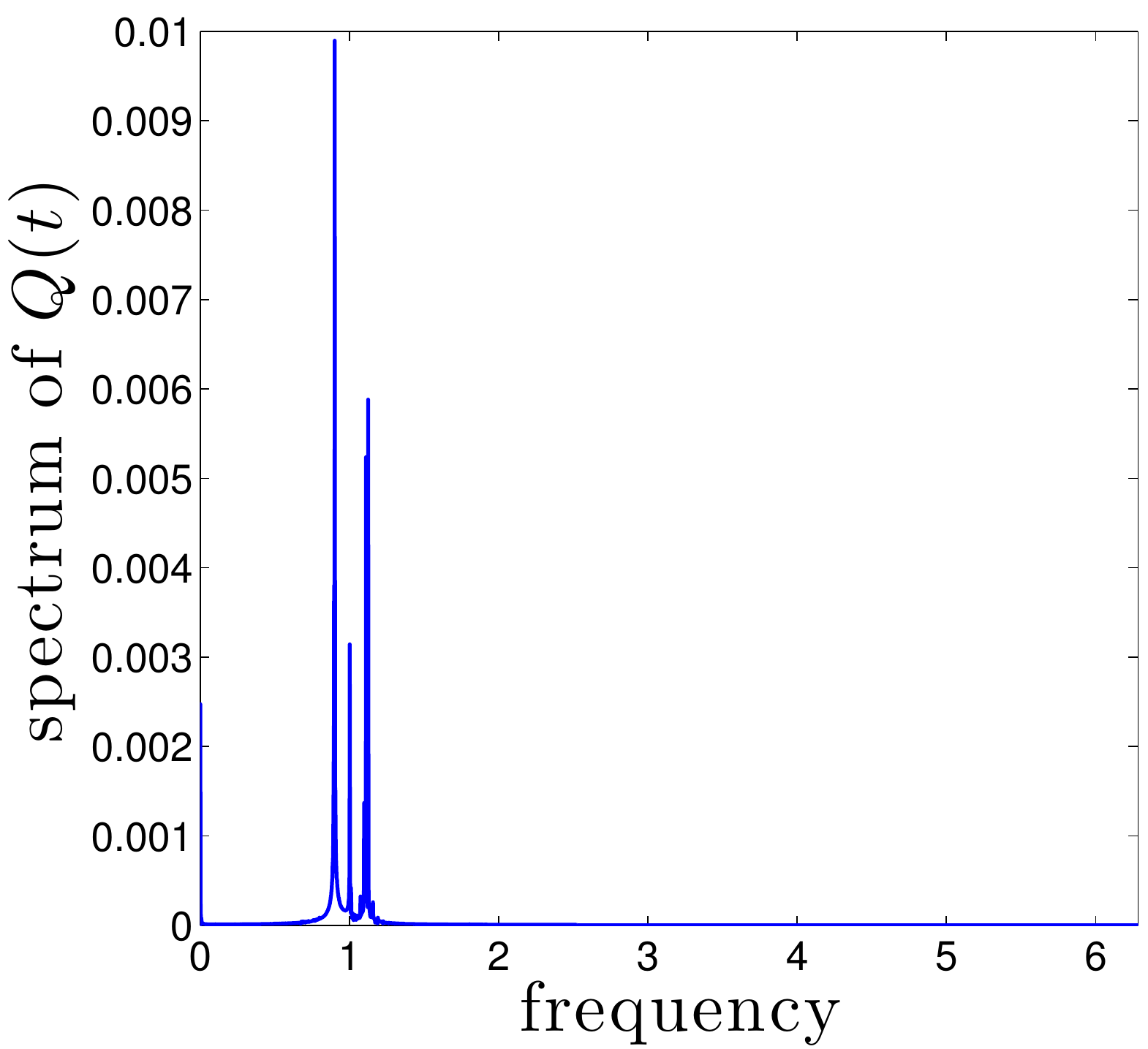} \\
 \\
\includegraphics[width=7.0cm]{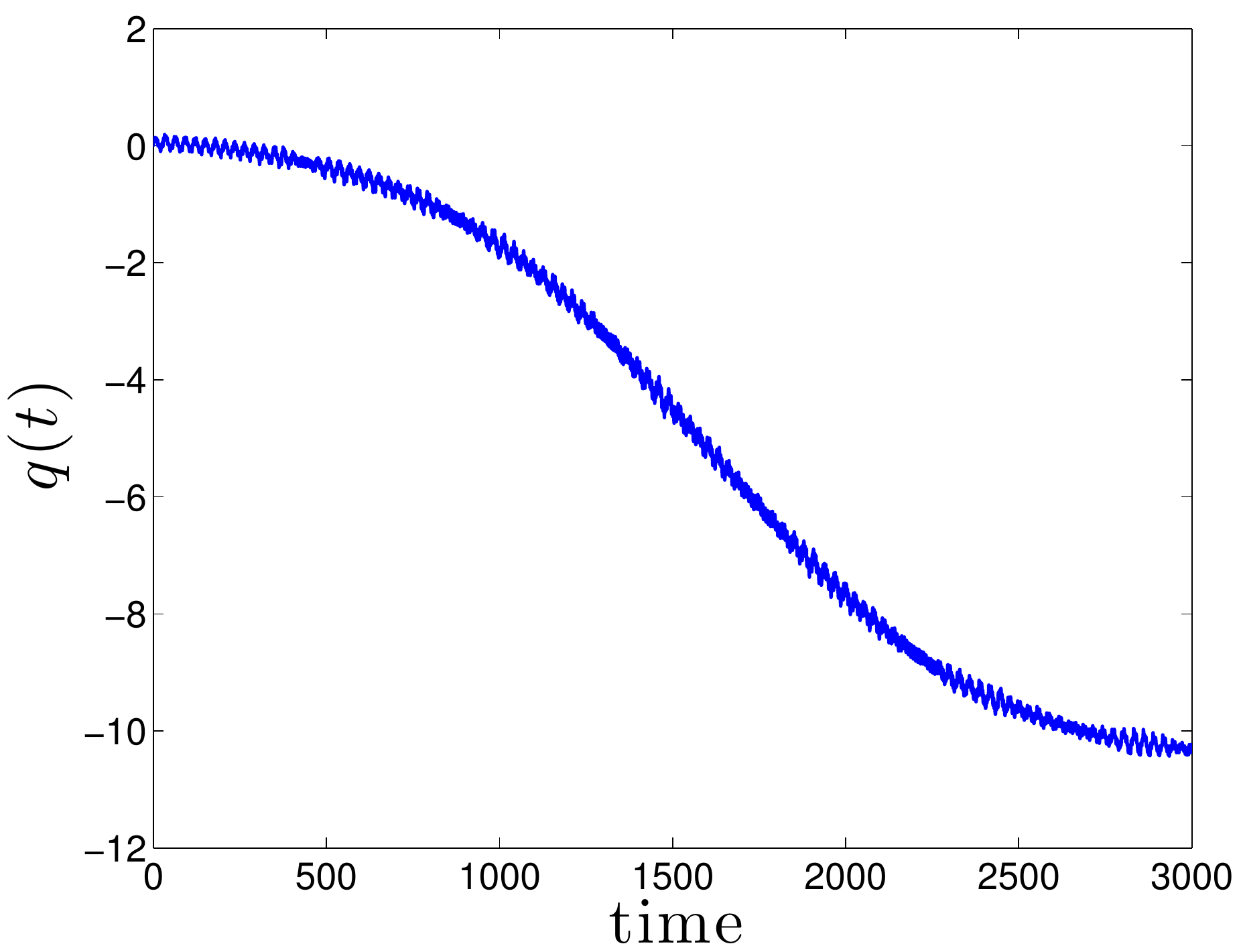}  & 
\quad \includegraphics[width=7.0cm]{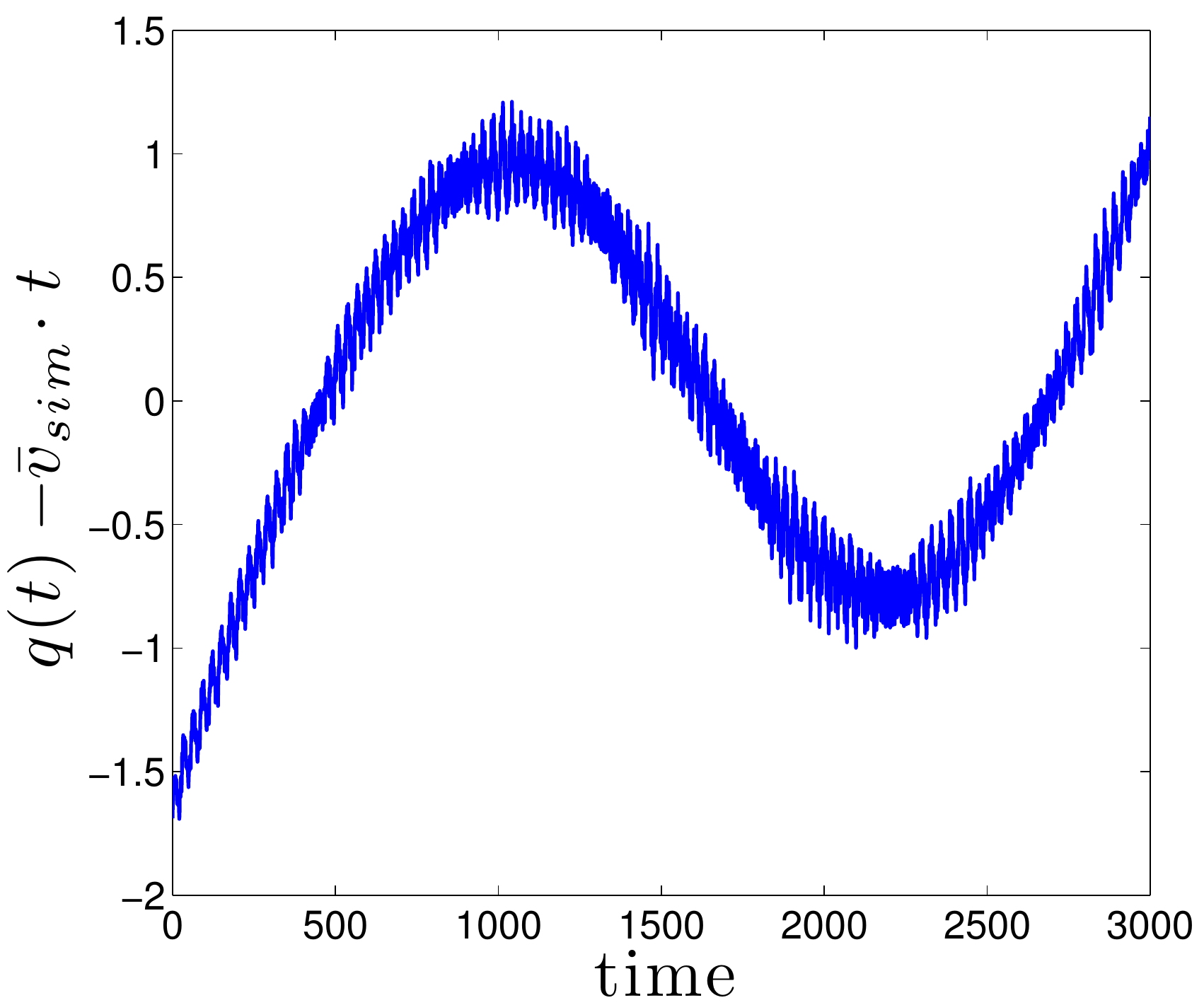}
\\ 
\includegraphics[width=7.0cm]{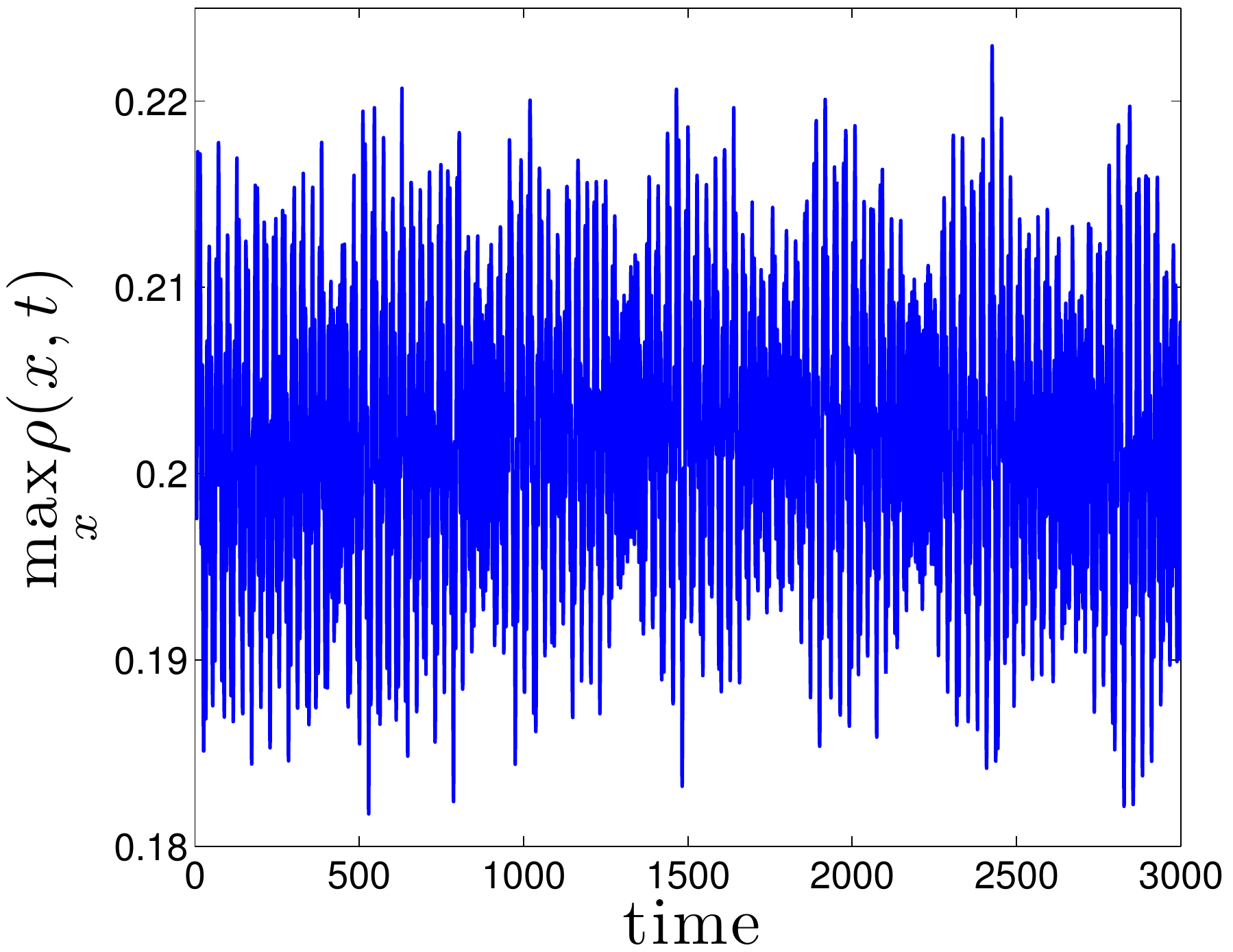}  & 
\quad \includegraphics[width=7.0cm]{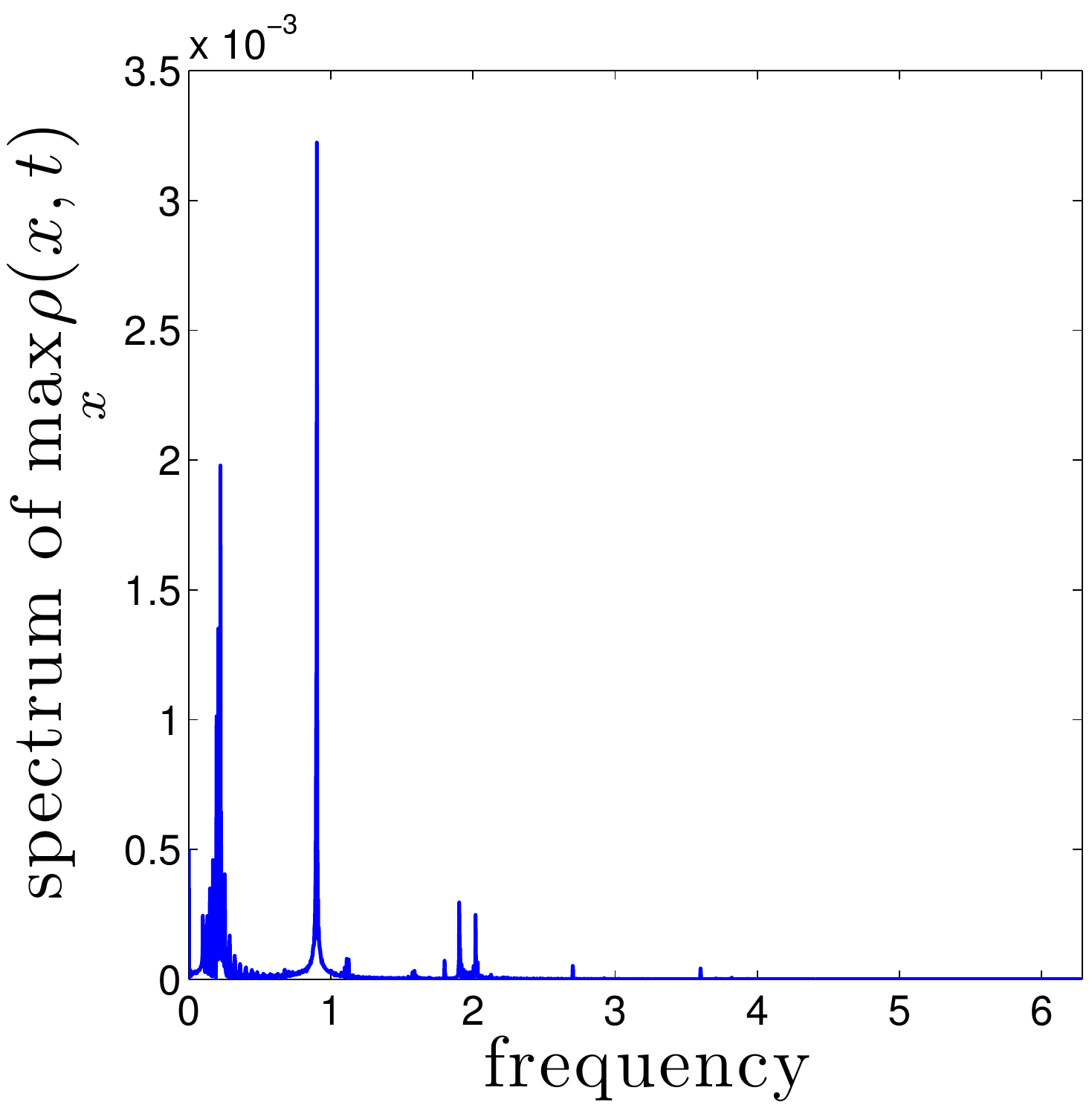}
\\
\end{tabular}
\end{center}
\caption{Simulation results: Oscillations of a traveling soliton. Same parameters and initial conditions as in Fig. 1.  
Left upper panel: Charge  $Q(t)$.  Right upper panel spectrum of $Q(t)$ with peaks at the frequencies $0.9027, 1.1268, 1.0032$ and $0.0020944 $.  Left middle panel: position $q(t)$.  Right middle panel:  $q(t)- \bar{v} t$.   
Left lower panel:  $max_x \rho(x,t)$.  Right lower panel:  spectrum of $max_x  \rho(x,t)$ with  peaks at $0.9027,  0.2241$, and $0.0020946.$
}
\label{fig4} 
\end{figure}


\section{Summary} \label{sec8}

We investigated the nonlinear Dirac (NLD) equation with an external spinor force with the components $ f_j = r_j exp(-i K_j x), j = 1,2$. In a previous paper (I) \cite{us1} we restricted ourselves to the case $K_1 = K_2,$ because in this case we could make a transformation on the wave function such that the Lagrangian was invariant under spatial translations, leading to a conserved momentum. Without this conservation law our collective coordinate (CC) theory became considerably more complicated: we had to solve a second-order ODE and two first-order ODEs, whereas in (I) we only had to solve two first-order ODEs and an algebraic equation. 
As an ansatz for our CC theory we took the exact Lorentz boosted solitary wave solution of the unperturbed NLD equation. The collective variables we chose are the soliton position $q(t)$, inverse width $\beta(t)$ and phase $\phi(t).$ The variable $\beta$ is related to the frequency $\omega(t) = \sqrt{m^2-\beta^2}$ that appears in the solitary wave solution and lies in the range $0 < \omega < m$. We restricted ourselves to the non-relativistic regime where $\omega$ is close to the mass $m$. Our ODEs for the CCs were solved numerically by a MATHEMATICA program. 

The solutions for all CCs are periodic in time, which means that the solitary waves exhibit intrinsic oscillations with a frequency $\omega^{cc}_s$. The translational motion of the soliton is also affected, but much stronger than in (I). There are two scenarios: In the first one the soliton is trapped and performs oscillations with a very low frequency $\omega^{cc}_q$, which is two orders of magnitude smaller than $\omega^{cc}_s$. The amplitude of these oscillations is much larger than the amplitude of the intrinsic oscillations. In the second scenario the soliton travels and performs very slow oscillations around a mean trajectory, again the amplitude is relatively large.
We compared our CC predictions with numerical simulations of the forced NLD equation. The solitary wave solutions are in fact stable, even for very long integration times. The observed frequency $\omega_s$ in the spectra of the charge, the amplitude, and the position is nearly identical with $\omega^{cc}_s$. Close to $ \omega_s $ there are two additional peaks in the spectrum of the charge which can be identified with two specific plane wave phonon modes which are excited together with the intrinsic oscillations. Moreover, the predicted scenarios of trapped and traveling solitons are observed and exhibit indeed very slow oscillations. However, their frequencies are considerably lower than $\omega^{cc}_q$, so there is only a qualitative agreement.
For the future we plan to work in the relativistic regime, i.e. $\omega_0$ not close to the mass $m$. Moreover, we want to consider the influence of time dependent external forces, i.e. non-vanishing $\nu_j$.

\section{Acknowledgments}  
This work was performed in part under the auspices of the United States Department of Energy. The authors would like to thank the Santa Fe Institute for its hospitality during the completion of this work. 
S.S. acknowledges financial support from the National Natural Science
Foundation of China (Nos.~11471025, 91330110, 11421101).
N.R.Q. 
acknowledges financial support from the Alexander von Humboldt Foundation (Germany) through Research Fellowship for Experienced Researchers SPA 1146358 STP and by the MICINN (Spain) through 
FIS2011-24540, and by Junta de Andalucia (Spain) under Projects No. FQM207, No. 
P06-FQM-01735, and No. P09-FQM-4643.  F.G.M. acknowledges the 
hospitality of the Mathematical Institute of the University of Seville (IMUS) and of the Theoretical 
Division and Center for Nonlinear Studies at Los Alamos National Laboratory, financial 
support by the Plan Propio of the University of Seville, and by the MICINN (Spain) through FIS2011-24540. 

\appendix
\section{Relevant Integrals} \label{sec9}

For our ansatz in the rest frame, we have that  for $\kappa=1$
%
the charge $Q$
is
\bq \label{A1}
 Q=\int dx \Psi^{\dag} \Psi= \int dx (A^2 +B^2)=\frac{2 \beta^2}{ g^2 (m+ \omega)}    \int_{- \infty}^\infty dx  \frac{1+ \alpha^2 \tanh^2 \beta x}{(1-\alpha^2 \tanh^2 \beta x)^2} \sech^2 \beta x
=  \frac{2 \beta}{g^2 \omega} . 
\eq
For Sec. V we need explicit expressions for the following  integrals  (in what follows,  $y= \tanh  \beta x$):
\ba
H_1 &&= -\frac{i}{2} \int dx \left[\bar \Psi \gamma^1 \partial_x \Psi -\partial_x \bar \Psi \gamma^1 \Psi \right] = \int dx  (B' A -A'B) = \frac{2 (m-\omega)}{g^2} \alpha  \int_{-1}^1 dy  \frac { 1-y^2} { ( 1- \alpha^2 y^2)^2} \nonumber \\
&&= \frac{2}{g^2} \left(2 \tanh ^{-1}\left(\sqrt{\frac{m-\omega }{\omega
   +m}}\right)-\beta \right) = I_0, \label{A2}
\ea
\ba
H_2 &=& m\int dx \bar \Psi \Psi =m I_1 = m \int dx (A^2 -B^2) =\frac{ 2 m \beta}{ g^2 (m+ \omega)}   \int_{-1}^1 dy  \frac { 1} { ( 1- \alpha^2 y^2)}= \frac{4 m \beta}{g^2 (m+ \omega)}  \frac{ \tanh ^{-1}(\alpha )}{\alpha } \nonumber \\
&& = \frac{4 m}{g^2} \tanh ^{-1}(\alpha )=M_0, \label{A3}
\ea
where $M_0$ is the mass in the rest frame. 
Note that $M_0$ has the property of vanishing as $\omega \to 1$.
\ba
I_2 && = \int dx (A^2 -B^2)^2 =\frac{ 4 \beta^3}{ g^4 (m+ \omega)^2}   \int_{-1}^1 dy  \frac { 1-y^2} { ( 1- \alpha^2 y^2)^2}= \frac{ 4 \beta^3}{ g^4 (m+ \omega)^2}  \left(\frac{\left(\alpha ^2+1\right) \tanh ^{-1}(\alpha )-\alpha }{\alpha ^3}\right) \nonumber \\
&& = \frac{2} {g^2} I_0=\frac{2}{g^2} H_1.  \label{A4}
 \ea
 To calculate the integral $J_j$ defined in (\ref{4.15}), first we rewrite it as 
\ba
J_j(\omega, \dot{q}) &=& \int_{-\infty}^{+\infty} dz A(z)  \cos(2 \beta a_j z) = 
 \frac{\sqrt{2 (m+\omega)} \beta}{g \omega}  \int_{-\infty}^{+\infty} dz   \frac{\cosh(\beta z) 
 \cosh(i 2 \beta a_j z)}{\frac{m}{\omega}+\cosh(2 \beta z)}  \nonumber \\
&=& \frac{\sqrt{2 (m+\omega)} \beta}{g \omega}  \int_{0}^{+\infty} dz   
\frac{\cosh[(1+i 2 a_j) \beta z]+ 
 \cosh[(1-i 2 a_j) \beta z]}{\frac{m}{\omega}+\cosh(2 \beta z)}.\label{A5}
 \ea
Now using expression (6) on page 357 of \cite{bk:PrudnikovBrychkov1986}, after some straightforward calculations we obtain
\ba  
J_j(\omega,\dot{q})&=&\frac{\pi \cos b_j}{g \sqrt{\omega} \cosh a_j \pi},  \label{A6}
\ea
where $a_j$ and $b_j$ are defined in Eq.\ (\ref{4.15}). The integral $N_j$ can be calculated in a similar way. 


%

\end{document}